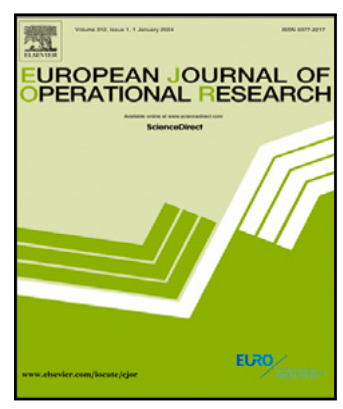

Discrete optimization

# Solving the Multiobjective Quasi-clique Problem

Daniela Scherer dos Santos [a],*, Kathrin Klamroth [b], Pedro Martins [c], Luís Paquete [a]

[a] CISUC, Department of Informatics Engineering, University of Coimbra, 3030-290, Portugal
[b] School of Mathematics and Natural Sciences, University of Wuppertal, 42119, Germany
[c] Coimbra Business School - ISCAC, Polytechnic Institute of Coimbra, 3045-601, Portugal

ARTICLE INFO

ABSTRACT

Given a simple undirected graph $G$, a quasi-clique is a subgraph of $G$ whose density is at least $\gamma$ $(0 < \gamma \leq 1)$. Finding a maximum quasi-clique has been addressed from two different perspectives: $(i)$ maximizing vertex cardinality for a given edge density; and $(ii)$ maximizing edge density for a given vertex cardinality. However, when no a priori preference information about cardinality and density is available, a more natural approach is to consider the problem from a multiobjective perspective. We introduce the *Multiobjective Quasi-clique* (MOQC) problem, which aims to find a quasi-clique by simultaneously maximizing both vertex cardinality and edge density. To efficiently address this problem, we explore the relationship among MOQC, its single-objective counterpart problems, and a bi-objective optimization problem, along with several properties of the MOQC problem and quasi-cliques. We propose a baseline approach using $\varepsilon$-constraint scalarization and introduce a *Two-phase* strategy, which applies a dichotomic search based on weighted sum scalarization in the first phase and an $\varepsilon$-constraint methodology in the second phase. Additionally, we present a *Three-phase* strategy that combines the dichotomic search used in *Two-phase* with a vertex-degree-based local search employing novel sufficient conditions to assess quasi-clique efficiency, followed by an $\varepsilon$-constraint in a final stage. Experimental results on synthetic and real-world sparse graphs indicate that the integrated use of dichotomic search and local search, together with mechanisms to assess quasi-clique efficiency, makes the *Three-phase* strategy an effective approach for solving the MOQC problem in sparse graphs in terms of running time and ability to produce new efficient quasi-cliques.

## 1. Introduction

Given a simple undirected graph $G = (V, E)$ with a set of vertices $V$ and a set of edges $E$, a subset of vertices $S \subseteq V$ is called a *quasi-clique* if the density of the subgraph $G_S$ induced by $S$ is at least a given threshold $\gamma \in (0, 1]$ (Abello et al., 1999, 2002).

One can define two optimization problem variants for quasi-cliques (Abello et al., 1999; Balasundaram & Pajouh, 2013): $(i)$ given a graph and $\gamma$, the objective is to discover a $\gamma$-quasi-clique with the maximum number of vertices; $(ii)$ given a graph and an integer constant $k$, the objective is to identify a subgraph of $k$ vertices with the maximum density (or, equivalently, the maximum edge cardinality). The variant $(i)$ is known as the *Maximum Quasi-Clique* (MQC) Problem (Abello et al., 1999) while $(ii)$ is commonly referred to under different terms in the literature, such as *Maximum Edge Subgraph Problem* (Asahiro et al., 2000), *Densest k-Set Problem* (Chang et al., 2014), *k-Cluster Problem* (Corneil & Perl, 1984), *Heaviest Unweighted Subgraph Problem* (Kortsarz & Peleg, 1993); in this work, we refer to $(ii)$ as the *Densest k-Subgraph* (DKS) Problem (Feige & Seltser, 1997). Both MQC and DKS problems are known to be NP-hard (Asahiro & Iwama, 1995; Feige & Seltser, 1997; Pattillo et al., 2013) and find applications in many real-world scenarios such as social networks (Brunato et al., 2008), telecommunications (Abello et al., 2002), and Bioinformatics (Althaus et al., 2014; Backes et al., 2011; Bhattacharyya & Bandyopadhyay, 2009). Consequently, many exact (Althaus et al., 2014; Billionnet, 2005; Bourgeois et al., 2013; Chang et al., 2014; Komusiewicz et al., 2015; Marinelli et al., 2021; Pajouh et al., 2014; Pattillo et al., 2013; Ribeiro & Riveaux, 2019; Veremyev et al., 2016), as well as heuristic approaches (Abello et al., 1999, 2002; Bhattacharyya & Bandyopadhyay, 2009; Chen et al., 2021; Djeddi et al., 2019; Macambira, 2002; Peng et al., 2021; Pinto et al., 2021, 2018; Zhou et al., 2020) have been developed to address these problems. Additionally, approximation algorithms have been proposed for addressing the DKS problem (Bhaskara et al., 2010; Bourgeois et al., 2013; Chen et al., 2017; Feige et al., 2001; Kortsarz & Peleg, 1993; Liazi et al., 2008).

* Corresponding author.
*E-mail addresses:* dssantos@dei.uc.pt (D.S. dos Santos), klamroth@uni-wuppertal.de (K. Klamroth), pmartins@iscac.pt (P. Martins), paquete@dei.uc.pt (L. Paquete).

Solving these problems requires providing a priori information, such as the minimum density threshold ($\gamma$) for MQC, and the desired vertex cardinality ($k$) for DKS. However, when considering these problems in the context of practical applications, specifying this information with a high degree of precision may be challenging and limiting. Fixing a constraint value, whether for cardinality or density leads to the inevitable loss of information about other potential solutions that remain unexplored. For instance, violating the vertex cardinality constraint may be acceptable if the trade-off concerning the potential improvement in density is favourable. Therefore, a possible way of addressing this problem is to consider a *multiobjective* perspective where density and cardinality are objectives to be maximized simultaneously. We call such a problem a *Multiobjective Quasi-clique* (MOQC) problem.

To the best of our knowledge, the MOQC problem has been only briefly mentioned in Saban et al. (2010) in the context of social network analysis of bilateral investment treaties among countries. This allows them to understand the formation of highly cohesive subgroups of countries, which are represented as quasi-cliques. However, the authors do not give any implementation details about the solution method. Additionally, we can consider approaches proposed for a related problem called the multiobjective subgraph mining problem (Shelokar et al., 2010, 2011, 2013a, 2013b, 2014). This problem involves mining subgraphs in a given set of graphs according to two or more objectives, such as subgraph frequency, number of vertices, density, connectivity, and diameter. However, the existing approaches to this problem are heuristic methods, whereas our work uniquely focuses on exact strategies.

In this article, we develop solution approaches to the MOQC problem. We explore the fact that an optimal solution to the DKS problem or the MQC problem is always a (weakly) efficient solution to the related MOQC problem, which naturally leads to an $\varepsilon$-*constraint*-type solution approach (Haimes et al., 1971) to the latter. Moreover, we establish a link between the set of efficient solutions to the MOQC problem and a subset of weakly-efficient solutions of a bi-objective optimization problem, for which it is possible to find a certain subset of efficient solutions in polynomial time. Additionally, we explore a particular property of quasi-cliques, known as *quasi-heredity* (Kosub, 2005; Pattillo et al., 2013), which allows to derive a local search approach to the new bi-objective problem with a guarantee of optimality under particular conditions. We investigate the combination of these techniques to develop an effective solution approach to the MOQC problem in sparse graphs and report experimental results on a wide set of benchmark instances.

The remainder of this work is organized as follows. Section 2 gives the relevant notations and definitions for the scope of this work. In Section 3 the main properties of the MOQC problem are presented. Section 4 presents our proposed strategies to approach the MOQC problem. Section 5 shows the computational experiments and results. Finally, the conclusions are given in Section 6.

## 2. Definitions and notations

In this section, we establish key notations and definitions used throughout this paper. We consider an undirected and simple graph $G = (V, E)$, where $V$ and $E$ are the vertex and edge sets of $G$, respectively. For a set of vertices $S \subseteq V$, we denote by $G_S = (S, E(S))$ the subgraph induced by $S$ in $G$. The *density* of $G_S$, denoted by $dens(G_S)$, is the ratio between the number of edges in $G_S$ and the number of edges in a complete graph with $|S|$ vertices, that is,

$$dens(G_S) = \frac{2 \cdot |E(S)|}{|S| \cdot (|S| - 1)}$$

The degree of a vertex $v$ in $G_S$ is the number of vertices in $G_S$ adjacent to $v$, and it is denoted by $deg_{G_S}(v)$. The minimum and the maximum degree of $G_S$ is denoted by $\delta(G_S)$ and $\Delta(G_S)$, respectively. The induced subgraph $G_S$ is called *clique* if $G_S$ is complete, that is, every two distinct vertices in $G_S$ are adjacent. The largest clique in $G$ is termed the *maximum clique*, and its size is denoted by $\omega(G)$. The problem of finding the maximum clique is known as the *Maximum Clique* problem.

In the following, we introduce the Maximum Quasi-Clique (MQC) problem and the Densest $k$-subgraph (DKS) problem.

**Definition 1** (*MQC Problem*)**.** Given a graph $G = (V, E)$ and a constant $\gamma$, where $0 < \gamma \leq 1$, find a subgraph $G_S$ induced by $S \subseteq V$ such that

$$S \in \arg\max_{S' \subseteq V} \left\{ |S'| \, : \, dens(G_{S'}) \geq \gamma \right\}$$

For $\gamma = 1$, the MQC problem becomes the Maximum Clique problem.

**Definition 2** (*DKS Problem*)**.** Given a graph $G = (V, E)$ and a positive integer $2 \leq k \leq |V|$, find a subgraph $G_S$ induced by $S \subseteq V$ such that

$$S \in \arg\max_{S' \subseteq V} \left\{ dens(G_{S'}) \, : \, |S'| = k \right\}$$

An equivalent formulation of the DKS problem is to consider maximizing the number of edges in the quasi-clique instead of its density. In this article, we refer to this as the *edge*-based DKS problem ($e$-DKS problem).

For a given graph $G$, we will use MQC($G$, $\gamma$), DKS($G$, $k$), and $e$-DKS($G$, $k$) when referring to any state-of-the-art approach that returns the optimal value for the MQC, DKS, and $e$-DKS problems for given parameters $\gamma$ and $k$, respectively. We now introduce the Multiobjective Quasi-Clique (MOQC) problem.

**Definition 3** (*MOQC Problem*)**.** Given a graph $G = (V, E)$, find a subgraph $G_S$ induced by $S \subseteq V$ such that

$$S \in \arg\max_{S' \subseteq V} \left\{ \left( dens(G_{S'}), |S'| \right) \right\}$$

We will use the following convention throughout the paper: Given that a subgraph of $k$ vertices with density $\gamma^*$ in the MOQC problem is a feasible $\gamma^*$-quasi-clique for the MQC problem with $\gamma = \gamma^*$, we state that any induced subgraph for the MOQC problem is also a quasi-clique.

The MOQC problem, in general, does not have a unique optimal quasi-clique because density and vertex cardinality are conflicting objectives. The largest quasi-clique is not necessarily the densest, and simultaneously, the densest is not necessarily the largest. This implies that improving one objective may lead to degradation in the other. Consequently, to assess the quality of the quasi-cliques, it is necessary to specify the notion of optimality to be used. In the following, we adopt the usual notion of *efficiency* in multiobjective optimization (Ehrgott, 2005).

A quasi-clique $G_S$, $S \subseteq V$, is *weakly-efficient* if there exists no other quasi-clique $G_{S'}$, for any $S' \subseteq V$, such that $dens(G_{S'}) > dens(G_S)$ and $|S'| > |S|$. If $G_S$ is weakly-efficient, then $z = (dens(G_S), |S|)$ is a *weakly-nondominated* point. A quasi-clique $G_S$ is called *efficient* if there exists no other quasi-clique $G_{S'}$ such that $dens(G_{S'}) \geq dens(G_S)$ and $|S'| \geq |S|$ with at least one strict inequality. If $G_S$ is efficient, then $z = (dens(G_S), |S|)$ is a *nondominated* point. The set of all weakly-efficient quasi-cliques is the *weakly-efficient set*, $\mathcal{E}^w_{\text{MOQC}}$, and the set of all weakly-nondominated points is the *weakly-nondominated set*, $\mathcal{Z}^w_{\text{MOQC}}$. The set of all efficient quasi-cliques is the *efficient set*, $\mathcal{E}_{\text{MOQC}}$, and the set of all nondominated points is the *nondominated set*, $\mathcal{Z}_{\text{MOQC}}$. Note that $\mathcal{E}_{\text{MOQC}} \subseteq \mathcal{E}^w_{\text{MOQC}}$ and $\mathcal{Z}_{\text{MOQC}} \subseteq \mathcal{Z}^w_{\text{MOQC}}$.

Let $\mathbb{R}^2_{\leq} = \{z \in \mathbb{R}^2 \, : \, z \leq \mathbf{0}\}$ be the negative orthant (that defines the dominating cone), and let $\mathcal{Z}^{\leq}_{\text{MOQC}} = conv\{\mathcal{Z}_{\text{MOQC}} \oplus \mathbb{R}^2_{\leq}\}$, where $\oplus$ represents the Minkowski sum. The boundary and interior of $\mathcal{Z}^{\leq}_{\text{MOQC}}$ are denoted by $bd(\mathcal{Z}^{\leq}_{\text{MOQC}})$ and $int(\mathcal{Z}^{\leq}_{\text{MOQC}})$, respectively. A nondominated point $z^* \in \mathcal{Z}_{\text{MOQC}}$ is *supported* if $z^* \in bd(\mathcal{Z}^{\leq}_{\text{MOQC}})$ and *nonsupported* if $z^* \in int(\mathcal{Z}^{\leq}_{\text{MOQC}})$. Moreover, if $z^*$ is supported and an extreme point of $\mathcal{Z}^{\leq}_{\text{MOQC}}$, then $z^*$ is an *extreme supported* point. The sets of all supported, all extreme supported, and all nonsupported points are denoted as

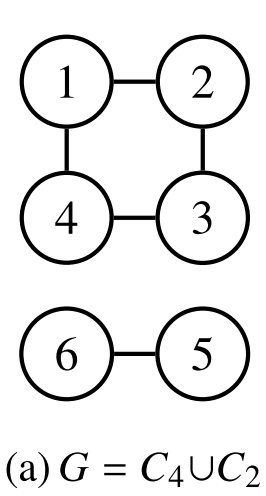

(a) $G = C_4 \cup C_2$

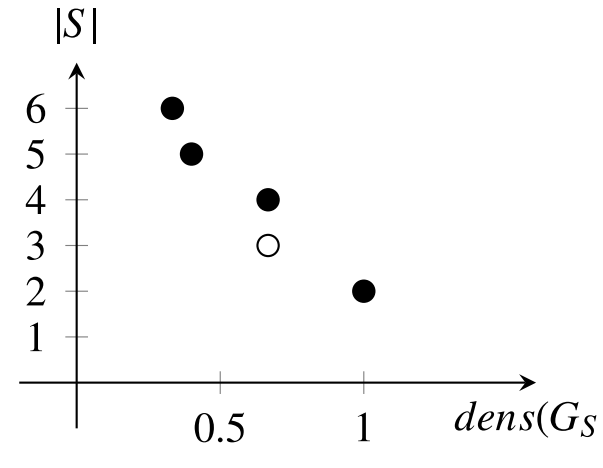

(b) Weakly-nondominated set $\mathcal{Z}^w_{\text{MOQC}}$ (all circles) and nondominated set $\mathcal{Z}_{\text{MOQC}}$ (solid circles) wrt. MOQC

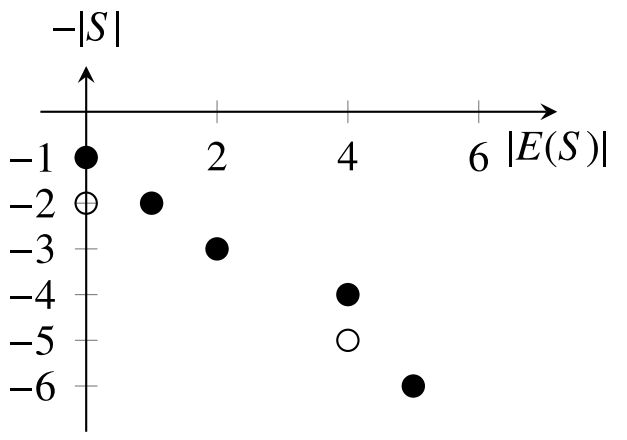

(c) Weakly-nondominated set $\mathcal{Z}^w_{\text{MOS}}$ (all circles) and nondominated set $\mathcal{Z}_{\text{MOS}}$ (solid circles) wrt. MOS

**Fig. 1.** Illustration of the (weakly) nondominated sets of problems MOQC and MOS, respectively, for the graph $G$ shown in (a).

$\mathcal{Z}^s_{\text{MOQC}}$, $\mathcal{Z}^e_{\text{MOQC}}$, and $\mathcal{Z}^n_{\text{MOQC}}$, respectively. The corresponding sets of quasi-cliques are denoted analogously as $\mathcal{E}^s_{\text{MOQC}}$, $\mathcal{E}^e_{\text{MOQC}}$, and $\mathcal{E}^n_{\text{MOQC}}$, respectively.

We also consider a related multiobjective optimization problem, the *Multiobjective Subgraph* (MOS) problem. This problem consists of finding a quasi-clique of $G$ that maximizes the number of edges and minimizes the number of vertices. Formally, it can be defined as follows:

**Definition 4** (*MOS Problem*)**.** Given a graph $G = (V, E)$, find a subgraph $G_S$ induced by $S \subseteq V$ such that

$$S \in \underset{S' \subseteq V}{\arg\max} \left\{ \left( |E(S')|, -|S'| \right) \right\}$$

The same notion of efficiency for the MOQC problem also applies to the MOS problem with the required changes. The weakly-efficient set of the MOS problem is denoted as $\mathcal{E}^w_{\text{MOS}}$, the weakly-nondominated set as $\mathcal{Z}^w_{\text{MOS}}$, the efficient set as $\mathcal{E}_{\text{MOS}}$, and the nondominated set as $\mathcal{Z}_{\text{MOS}}$. The remaining sets of points and solutions are denoted analogously.

Furthermore, similar to the assertion made for the MOQC problem, we state that any induced subgraph in the MOS problem is also a quasi-clique.

Fig. 1 illustrates the (weakly) nondominated sets of problems MOQC and MOS for the graph $G = C_4 \cup C_2$ shown in Fig. 1(a). Note that while the point $(0, -1)$ (corresponding to an induced subgraph with only one node) is always nondominated for problem MOS, a corresponding point does not exist for problem MOQC since the density is only defined for induced subgraphs with at least two nodes.

## 3. Main properties of the MOQC problem

In this section, we enumerate some properties of the MOQC problem which allows us to derive the main algorithmic results in later sections.

### 3.1. Basic properties

We first state the main results in terms of (in)tractability and monotonicity.

**Proposition 1.** *For a given graph $G = (V, E)$, $|\mathcal{Z}_{MOQC}| \le |V| - 1$.*

**Proof.** This follows from the fact that the second objective function of problem MOQC can have at most $|V| - 1$ different values, and each such value can occur in at most one nondominated point. □

The property stated in Proposition 1, which does not hold in general for multiobjective combinatorial optimization problems (see, e.g., the discussion in Figueira et al. (2017)), makes the MOQC problem particularly appealing to $\varepsilon$-constraint methods (Haimes et al., 1971). However, despite a tractable number of nondominated points, finding a quasi-clique by $\varepsilon$-constraint scalarization is an NP-hard problem since it implies solving either a DKS or an MQC problem. Note, however, that some nondominated points can be found easily, for instance, graph $G$ is always an efficient quasi-clique.

**Proposition 2.** *For a given graph $G = (V, E)$ and a given parameter $k$, $2 < k \le |V|$, let the quasi-cliques $G_{S^{k-1}}$ and $G_{S^k}$ be subgraphs induced by $S^{k-1} \subseteq V$ and $S^k \subseteq V$, respectively, such that*

$$S^{k-1} \in \underset{S' \subseteq V}{\arg\max} \left\{ dens(G_{S'}) : |S'| = k - 1 \right\}$$

$$S^{k} \in \underset{S' \subseteq V}{\arg\max} \left\{ dens(G_{S'}) : |S'| = k \right\}$$

*Then, $dens(G_{S^k}) \le dens(G_{S^{k-1}})$.*

**Proof.** Let $v \in S^k$ be a vertex of minimum degree, $deg_{G_{S^k}}(v) = \delta(G_{S^k})$ in $G_{S^k}$, i.e., $deg_{G_{S^k}}(v) \le \frac{2\cdot|E(S^k)|}{k}$. Now let $G_{S'}$ be a subgraph induced by $S' = S^{k-1} \setminus \{v\}$. Then, $dens(G_{S^{k-1}}) \ge dens(G_{S'})$ by definition of $G_{S^{k-1}}$. Moreover, $dens(G_{S'}) = \frac{2(|E(S^k)| - deg_{G_{S^k}}(v))}{(k-1)(k-2)} \ge \frac{2(|E(S^k)| - 2|E(S^k)|/k)}{(k-1)(k-2)} = \frac{2|E(S^k)|\cdot(k-2)/k}{(k-1)(k-2)} = dens(G_{S^k})$, which completes the proof. □

From the monotonicity property in Proposition 2, we can derive that the quasi-clique with maximum density for a given fixed cardinality is always weakly-efficient. This will be analysed in more detail in Section 3.2 below. Furthermore, this implies that the largest quasi-clique with a density larger than a given $\gamma$ is also weakly-efficient.

Finally, note that, for a given graph $G = (V, E)$, the set $\mathcal{E}_{\text{MOQC}}$ contains the following two (lexicographically optimal) quasi-cliques: $(i)$ the maximum clique of $G$, and $(ii)$ the graph $G$.

### 3.2. Relation between DKS, MQC, and MOQC

Addressing a constrained optimization problem as a multiobjective optimization problem has been discussed in the literature. For instance, Klamroth and Tind (2007) establishes interrelations between both problems, which allow to develop general methods that can solve the multiobjective optimization problem by solving associated constrained optimization problems and vice versa. In fact, $\varepsilon$-constraint methods (Haimes et al., 1971) explore this intertwining between the two types of problems. In the following, we enumerate some properties that relate MOQC with the constrained versions, DKS and MQC.

**Proposition 3.** *For a given graph $G = (V, E)$, and given parameters $k$, $1 < k \le |V|$, and $\gamma \in (0, 1]$, let the quasi-cliques $G_{S^k}$ and $G_{S^\gamma}$ be subgraphs induced by $S^k \subseteq V$ and $S^\gamma \subseteq V$, respectively, such that*

$$S^{k} \in \underset{S' \subseteq V}{\arg\max} \left\{ dens(G_{S'}) : |S'| = k \right\}$$

$$S^{\gamma} \in \underset{S' \subseteq V}{\arg\max} \left\{ |S'| : dens(G_{S'}) \ge \gamma \right\}$$

*Then, $\{G_{S^k}, G_{S^\gamma}\} \subseteq \mathcal{E}^w_{MOQC}$.*

**Proof.** By Proposition 2 and the definition of $G_{S^k}$, there is no subgraph $G_{S^{k+i}}$ with $k + i$ vertices such that $dens(G_{S^{k+i}}) > dens(G_{S^k})$, for $i = 1, \ldots, |V| - k$. Moreover, the definition of $G_{S^\gamma}$ implies that there is no subgraph $G_{S'}$ with $dens(G_{S'}) > \gamma$ and $|S'| > |S^\gamma|$. □

**Algorithm 1:** MQC-DKS-based $\varepsilon$-constraint method for problem MOQC

**Data:** $G = (V, E), e > 0$

**Result:** $\mathcal{Z}_{\text{MOQC}}$

1 $\mathcal{Z}_{\text{MOQC}} \leftarrow \{(dens(G), |V|)\}$
2 $\gamma \leftarrow dens(G)$
3 **while** $\gamma < 1$ **do**
4 $\quad k \leftarrow MQC(\gamma + e)$
5 $\quad \gamma \leftarrow DKS(k)$
6 $\quad \mathcal{Z}_{\text{MOQC}} \leftarrow \mathcal{Z}_{\text{MOQC}} \cup \{(\gamma, k)\}$
7 **end**

Proposition 3 states that an optimal quasi-clique for both the MQC problem and the DKS problem is weakly-efficient for the related MOQC problem. Therefore, solving the MQC problem for a range of $\gamma$ values enables the identification of the weakly-efficient set for the MOQC problem. The same principle applies to the DKS problem for $k = 1, \dots, |V|$. This forms the basis for developing the following two $\varepsilon$-constraint methods for MOQC.

Algorithm 1 shows the pseudo-code of the first $\varepsilon$-constraint method to obtain the nondominated set $\mathcal{Z}_{\text{MOQC}}$ of a MOQC problem, by an alternating sequence of MQC and DKS problems (see Lines 4 and 5, respectively), where at each iteration of the main loop, a nondominated point is found. The first nondominated point corresponds to the density and number of vertices of graph $G$ (Line 1). If $G$ is a complete graph, the algorithm terminates; otherwise, it enters the main loop. Subsequently, an MQC problem is solved for a given $\gamma + e$, where $e$ is a sufficiently small positive constant. Next, the cardinality $k$ of each weakly-efficient quasi-clique found for the MQC problem (Line 4) is used as a cardinality constraint for the related DKS problem (Line 5), which provides a nondominated point corresponding to the densest efficient quasi-clique with $k$ vertices. The algorithm terminates when a maximum clique of $G$ is found. Note that this approach requires at most $|\mathcal{Z}_{\text{MOQC}}|$ MQC problems and $|\mathcal{Z}_{\text{MOQC}}|$ DKS problems to be solved.

Algorithm 2 shows the pseudo-code of an $\varepsilon$-constraint method based only on solving a sequence of DKS problems. Given the equality constraint in the DKS problem, it is necessary to consider all possible cardinality values, ranging from $|V|$ to the size of the maximum clique. We recall from Proposition 3 that every optimal quasi-clique for the DKS problem is also weakly-efficient for MOQC. Therefore, using the main result from Proposition 2, for each quasi-clique of size $k$ that maximizes density, the algorithm verifies if the previously found efficient quasi-clique has a smaller density (Line 7). If so, the quasi-clique is efficient, and the corresponding nondominated point is stored. Note that this approach requires solving at most $|V|$ DKS problems.

It is worth mentioning that a similar strategy could be employed using the MQC problem with an equality constraint on the density value (an *MQC-based* variant). However, this would imply a discretization of all possible density values.

*3.3. Relation between MOQC and MOS problems*

In this section, we investigate the relation between the MOS problem and the MOQC problem.

**Proposition 4.** *For a given graph $G = (V, E)$,*

*(i) Let $G_S$ be an efficient quasi-clique for the MOQC problem. Then, $G_S$ is a weakly-efficient quasi-clique for the MOS problem;*
*(ii) Let $G_S$ be an efficient quasi-clique for the MOS problem with $|S| \geq 2$. Then, $G_S$ is a weakly-efficient quasi-clique for the MOQC problem.*

**Proof.** For a given cardinality $k$, $2 \leq k \leq |V|$, $G_S$ is maximal in terms of density and number of edges. □

**Algorithm 2:** DKS-based $\varepsilon$-constraint method for problem MOQC

**Data:** $G = (V, E)$

**Result:** $\mathcal{Z}_{\text{MOQC}}$

1 $\mathcal{Z}_G \leftarrow \{(dens(G), |V|)\}$
2 $k \leftarrow |V|$
3 $\gamma \leftarrow dens(G)$
4 **while** $\gamma < 1$ **do**
5 $\quad k \leftarrow k - 1$
6 $\quad \gamma' \leftarrow DKS(k)$
7 $\quad$ **if** $\gamma' > \gamma$ **then**
8 $\quad\quad \gamma \leftarrow \gamma'$
9 $\quad\quad \mathcal{Z}_{\text{MOQC}} \leftarrow \mathcal{Z}_{\text{MOQC}} \cup \{(\gamma, k)\}$
10 $\quad$ **end**
11 **end**

Proposition 4 suggests that it is possible to find the efficient set for the MOQC problem by enumerating all weakly-efficient quasi-cliques for the MOS problem. In the following, we show that collecting only efficient quasi-cliques for the MOS problem is not enough.

**Proposition 5.** *For a given graph $G = (V, E)$, let $G_S$ be an efficient quasi-clique for the MOQC problem. Then, $G_S$ is not necessarily efficient for the MOS problem.*

**Proof.** Assume for the sake of contradiction that $G_S$ is always efficient for the MOS problem if it is efficient for the MOQC problem. We demonstrate that this assumption is false by providing a counterexample. Consider the graph $G = (V, E)$ with vertex set $V = \{v_1, v_2, v_3, v_4, v_5, v_6\}$ and edge set $E = \big\{\{v_1, v_2\}, \{v_1, v_4\}, \{v_2, v_3\}, \{v_3, v_4\}, \{v_5, v_6\}\big\}$, as shown in Fig. 1. Let $S = \{v_1, v_2, v_3, v_4, v_5\}$. We calculate the number of edges $|E(S)|$ and the density $dens(G_S)$ in the subgraph $G_S$ induced by $S$: $|E(S)| = 4$ and $dens(G_S) = \frac{2 \cdot |E(S)|}{|S| \cdot (|S|-1)} = \frac{2}{5}$. The quasi-clique $G_S$ is efficient for the MOQC problem because there is no other quasi-clique $G_{S'}$ such that $|S'| \geq |S|$ and $dens(G'_S) \geq dens(G_S)$ with at least one strict inequality. Now, consider the set $S'' = \{v_1, v_2, v_3, v_4\}$. For $G_{S''}$ the number of edges $|E(S'')| = 4$. Therefore, $G_S$ is not efficient for the MOS problem because $G_{S''}$ dominates $G_S$ by having the same number of edges ($|E(S'')| = |E(S)|$) and fewer vertices ($|S''| < |S|$). □

In fact, not all weakly-efficient quasi-cliques need to be considered for the MOS problem, as shown in the following proposition.

**Proposition 6.** *For a given graph $G = (V, E)$, let the subgraphs $G_S$ and $G_{S'}$ induced by $S \subseteq V$ and $S' \subseteq V$, respectively, be weakly-efficient quasi-cliques for the MOS problem, such that $|E(S)| > |E(S')|$ and $|S| = |S'|$. Then, $G_{S'}$ is not efficient for the MOQC problem.*

**Proof.** This follows immediately from the fact that $|E(S)| > |E(S')|$ and $|S| = |S'|$ imply $dens(G_S) > dens(G_{S'})$. □

Note that, analogous to Proposition 3 we immediately have that an optimal solution of the $e$-DKS problem is at least weakly-efficient for problem MOS. Therefore, following Proposition 6, it is enough to consider the quasi-cliques that maximize the number of edges, considering a fixed cardinality constraint on the number of vertices, from $\omega(G)$ to $|V|$, i.e., solutions of problem $e$-DKS for $\omega(G) \leq k \leq |V|$. We denote the set of points of all such weakly-efficient quasi-cliques as $\mathcal{Z}^V_{\text{MOS}}$.

Algorithm 3 shows a pseudo-code to an $\varepsilon$-constraint approach that allows finding the set $\mathcal{Z}^V_{\text{MOS}}$ for the MOS problem for a given graph $G = (V, E)$. Similar to Algorithm 2, it collects the lexicographical optimal point with the number of edges and vertices of the graph $G$. Then, by decreasing values of cardinality $k$, it solves $e$-DKS problems

**Algorithm 3:** $e$-DKS-based $\varepsilon$-constraint method to find the set $\mathcal{Z}^V_{\text{MOS}}$ for MOS

**Data:** $G = (V, E)$
**Result:** $\mathcal{Z}^V_{\text{MOS}}$
1 $\mathcal{Z}^V_{\text{MOS}} \leftarrow \{(|E|, -|V|)\}$
2 $k \leftarrow |V|$
3 $m \leftarrow |E|$
4 **while** $2 \cdot m/(k^2 - k) < 1$ **and** $k > 1$ **do**
5 $\quad k \leftarrow k - 1$
6 $\quad m \leftarrow e\text{-}DKS(G, k)$
7 $\quad \mathcal{Z}^V_{\text{MOS}} \leftarrow \mathcal{Z}^V_{\text{MOS}} \cup \{(m, -k)\}$
8 **end**

**Algorithm 4:** dichotomicSearch

**Input:** $z^r$, $z^s$, $\mathcal{E}^e_{\text{MOS}}$, $\mathcal{Z}^e_{\text{MOS}}$, $G$
1 $w_1 \leftarrow z^r_2 - z^s_2$
2 $w_2 \leftarrow z^s_1 - z^r_1$
3 $G_S \leftarrow$ WS-MOS$(G, w_1, w_2)$
4 $z^t \leftarrow (|E(S)|, -|S|)$
5 **if** $z^t \neq z^r$ **and** $z^t \neq z^s$ **then**
6 $\quad \mathcal{E}^e_{\text{MOS}} \leftarrow \mathcal{E}^e_{\text{MOS}} \cup \{G_S\}$
7 $\quad \mathcal{Z}^e_{\text{MOS}} \leftarrow \mathcal{Z}^e_{\text{MOS}} \cup \{z^t\}$
8 $\quad \mathcal{E}^e_{\text{MOS}}, \mathcal{Z}^e_{\text{MOS}} \leftarrow dichotomicSearch(z^r, z^t, \mathcal{E}^e_{\text{MOS}}, \mathcal{Z}^e_{\text{MOS}}, G)$
9 $\quad \mathcal{E}^e_{\text{MOS}}, \mathcal{Z}^e_{\text{MOS}} \leftarrow dichotomicSearch(z^t, z^s, \mathcal{E}^e_{\text{MOS}}, \mathcal{Z}^e_{\text{MOS}}, G)$
10 **return** $\mathcal{E}^e_{MOS}, \mathcal{Z}^e_{MOS}$

until either a maximum clique is found (i.e., the density of the solution found is equal to 1) or the algorithm has solved for all values of $k$ ranging from $|V| - 1$ to 2.

From $\mathcal{Z}^V_{\text{MOS}}$, it is possible to extract the set of nondominated points $\mathcal{Z}_{\text{MOQC}}$ through a *mapping* process as follows. We compute the density of each point in $\mathcal{Z}^V_{\text{MOS}}$ and select all resulting nondominated points to the set $\mathcal{Z}_{\text{MOQC}}$. Recall that, according to Proposition 2, the density of a quasi-clique $G_S$ of size $k$ is always smaller than or equal to the density of a quasi-clique of size $k - 1$. Therefore, in the selecting step, weakly-nondominated points for MOQC in $\mathcal{Z}^V_{\text{MOS}}$ only arise for sequential values of $k$ with the same density value. In such cases, we select only the points with the highest value of $k$, ensuring that only nondominated points are included in the set $\mathcal{Z}_{\text{MOQC}}$.

### 3.4. Weighted-sum scalarization of the MOS problem

In this section, we show that extreme supported points of the MOS problem can be computed in polynomial time. To this end, we propose the following linear programming relaxation of the weighted-sum scalarization of the MOS problem (WS-MOS) for a given graph $G = (V, E)$.

$$\max \quad w_1 \cdot \sum_{\{i,j\} \in E} y_{ij} - w_2 \cdot \sum_{i \in V} x_i$$

$$\text{s. t.} \quad y_{ij} \leq x_i \qquad \forall \{i, j\} \in E \tag{1a}$$

$$y_{ij} \leq x_j \qquad \forall \{i, j\} \in E \tag{1b}$$

$$y_{ij} \geq 0 \qquad \forall \{i, j\} \in E \tag{1c}$$

$$x_i \geq 0 \qquad \forall i \in V \tag{1d}$$

$$x_i \leq 1 \qquad \forall i \in V \tag{1e}$$

The variables $x_i$ and $y_{ij}$ are defined for each vertex $i \in V$ and for each edge $\{i, j\} \in E$, respectively. Variable $x_i = 1$ indicates that vertex $i \in V$ is chosen, and 0 otherwise, and variable $y_{ij} = 1$ indicates that the edge connecting vertices $i$ and $j$ is chosen, 0 otherwise, and $w_1, w_2 > 0$ are the weights assigned to optimizing the number of edges and vertices, respectively. Constraints (1a) and (1b) state that if an edge $\{i, j\}$ is in the quasi-clique, then both vertices $i$ and $j$ must be chosen. Note that the constraint matrix corresponding to (1a), (1b), and (1e), is totally unimodular (Padberg, 1989). Therefore, this linear programming relaxation is integral.

For a given graph $G$, we will use WS-MOS$(G, w_1, w_2)$ when referring to a linear programming solver that returns the optimal value for the WS-MOS problem for parameters $w_1$ and $w_2$.

Optimal quasi-cliques to the WS-MOS formulation are also supported points for the MOS problem (Ehrgott, 2005). The set of extreme supported points, $\mathcal{Z}^e_{\text{MOS}}$, can be obtained by bisection methods, such as the dichotomic search based on weighted sum scalarization proposed by Aneja and Nair (1979). Algorithm 4 shows the pseudo-code of a recursive dichotomic search to find the set $\mathcal{Z}^e_{\text{MOS}}$, where $z^r$ and $z^s$ are two lexicographical optimal points with $z^r_1 \leq z^s_1$ and $z^r_2 \geq z^s_2$. The dichotomic search is initially triggered with $z^r = (0, 0)$, $z^s = (|E|, -|V|)$, and $\mathcal{Z}^e_{\text{MOS}} = \{z^r, z^s\}$. Note that $z^r_2 > z^s_2$ and $z^r_1 < z^s_1$ always holds during its run. Next, it computes an optimal point $z^t$ by solving WS-MOS (Line 3) with weights that are defined orthogonally to the vector between $z^r$ and $z^s$ (Lines 1 and 2). If $z^t$ is a new extreme supported point found between $y^r$ and $y^s$ (see condition in Line 5), it is added to the set $\mathcal{Z}^e_{\text{MOS}}$ and the resolution of two additional problems is triggered: one with weights defined by $z^r$ and $z^t$, and another with weights defined by $z^t$ and $z^s$ (Lines 8 and 9, respectively). Otherwise, no new point is found, and there is no need to further bisect that region. The procedure naturally terminates when no new extreme supported point is found, which means that the set $\mathcal{Z}^e_{\text{MOS}}$ has been found.

**Remark 1.** While it is arguable whether the vertex-minimizing extreme supported point should be $z^r = (0, 0)$ (an empty "quasi-clique" with no vertices) or $z^r = (0, -1)$ (the smallest possible subgraph with a non-empty set of vertices), we choose to initiate the dichotomic scheme with $z^r = (0, 0)$. Otherwise, i.e., when initiating the dichotomic scheme with $z^r = (0, -1)$, the constraint matrix of the WS-MOS model (1) needs to be extended by the constraint $\sum_{i \in V} x_i \geq 1$ which destroys the total unimodularity property, and hence makes the solution of the WS-MOS model (1) considerably more complex. In the following, by slightly abusing notation we refer to the set $\mathcal{Z}^e_{\text{MOS}}$ as the set of extreme supported nondominated points that is obtained when considering $z^r = (0, 0)$ and $z^s = (|E|, -|V|)$ as the two bounding outcome vectors corresponding to the individual maxima of $-|S|$ and $|E(S)|$, respectively. Note that this choice only increases the considered node interval and hence it does not hinder the identification of the complete nondominated set of problem MOS.

Since the number of extreme supported points is bounded by the number of vertices, then, the set $\mathcal{Z}^e_{\text{MOS}}$ can be found in a polynomial amount of time. Therefore, a subset of the weakly-nondominated points for MOQC problem can also be found in a polynomial amount of time.

### 3.5. Additional quasi-clique properties

**Proposition 7.** *For a given graph G = (V,E) and a given $\gamma$, let the subgraph $G_S$ induced by $S \subseteq V$ be a quasi-clique with $dens(G_S) = \gamma$. Let $v$ be a vertex in S with the smallest degree in $G_S$. Let $S' = S \setminus \{v\}$. Then, $dens(G_{S'}) \geq \gamma$.*

**Proof.** This is the quasi-heredity property defined in Kosub (2005) and Pattillo et al. (2013). □

**Proposition 8.** *For a given graph $G = (V, E)$, let the subgraph $G_S$ induced by $S \subseteq V$ be a quasi-clique. Let $v$ be a vertex in S with the smallest degree in $G_S$. Let $S' = S \cup \{v\}$. Then, $dens(G_S) \leq dens(G_{S'})$.*

**Proof.** This follows directly from Proposition 7. □

These propositions motivate a heuristic that, starting from an efficient quasi-clique for the MOS Problem, removes the vertex with the smallest degree. Although this selection represents the best local choice available, the quasi-clique obtained through this process may not necessarily be weakly-efficient for MOS. However, its (weak) efficiency can be evaluated using Propositions 9, 10, and 11, which are sufficient conditions and will be presented in Section 4.3.1.

Furthermore, Proposition 7 has already been explored in Komusiewicz et al. (2015), Ribeiro and Riveaux (2019), and Veremyev et al. (2016) to design exact approaches to the single-objective MQC problem (see Definition 1).

## 4. Proposed approaches

Given the relations between the MOQC and MOS problems (see Section 3.3 and the propositions therein), as well as the tractability of MOS in terms of computing extreme supported points (see Section 3.4), we focus on addressing the MOS problem, in order to solve MOQC more efficiently. The approaches proposed to achieve this goal are described in this section.

To begin, we introduce a basic $\varepsilon$-constraint approach, establishing it as a baseline for comparing subsequent strategies. Next, we present a *Two-phase* strategy that employs a dichotomic search based on weighted sum scalarizations to discover a set of extreme supported points, followed by an $\varepsilon$-constraint method to identify the remaining weakly-nondominated points. Following this, we propose a *Three-phase* strategy. In addition to a dichotomic search phase and an $\varepsilon$-constraint method, this strategy incorporates a local search technique grounded in vertex degree information. This local search aims to identify new quasi-cliques that are guaranteed to be efficient under specific conditions.

### 4.1. Baseline approach

The $e$-DKS-based $\varepsilon$-constraint method (Algorithm 3) is proposed to find the set $\mathcal{Z}^V_{\text{MOS}}$ of weakly-nondominated points for the MOS problem. For a given graph $G$, the algorithm iteratively solves each $e$-DKS scalarized problem. Here, $e$-DKS($G$, $k$) (Line 6) denotes the Mixed Integer Linear Programming (MILP) M1 model (Billionnet, 2005), as detailed below.

$$\max \sum_{\{i,j\}\in E} y_{ij}$$

$$\text{s. t.} \quad \sum_{i\in V} x_i = k \tag{2a}$$

$$y_{ij} \le x_i \qquad \forall \{i,j\} \in E \tag{2b}$$

$$y_{ij} \le x_j \qquad \forall \{i,j\} \in E \tag{2c}$$

$$y_{ij} \ge 0 \qquad \forall \{i,j\} \in E \tag{2d}$$

$$x_i \in \{0,1\} \qquad \forall\, i \in V \tag{2e}$$

The variables $x_i$ and $y_{ij}$ are defined as in the model WS-MOS (in Section 3.4), for each vertex $i \in V$ and for each edge $\{i,j\} \in E$, respectively. The objective function maximizes the number of edges of the quasi-clique. Constraint (2a) ensures that the cardinality of the quasi-clique is equal to $k$. Constraints (2b) and (2c) state that if an edge $\{i,j\}$ is in the quasi-clique, then both vertices $i$ and $j$ must be chosen.

This approach requires solving $|V| - \omega(G) + 1$ instances of the $e$-DKS problem, implying $e$-DKS($G$, $k$) to be executed $|V| - \omega(G) + 1$ times as well.

With the aim of reducing the number of runs of $e$-DKS($G$, $k$) and, as a result, decreasing the overall time required to solve the MOS problem, we introduce the *Two-phase* and *Three-phase* strategies, which are presented in the following sections.

**Algorithm 5:** Two-phase strategy for MOS

**Data:** $G = (V, E)$
**Result:** $\mathcal{Z}^V_{\text{MOS}}$ and $\mathcal{E}^w_{\text{MOS}}$

1 $z^r \leftarrow (0,0)$ // First phase
2 $z^s \leftarrow (|E|, -|V|)$
3 $\mathcal{Z}^e_{\text{MOS}} \leftarrow \{z^r, z^s\}$
4 $\mathcal{E}^e_{\text{MOS}} \leftarrow \{G_{S^r}, G\}$
5 $\mathcal{E}^e_{\text{MOS}}, \mathcal{Z}^e_{\text{MOS}} \leftarrow \mathit{dichotomicSearch}(z^r, z^s, \mathcal{E}^e_{\text{MOS}}, \mathcal{Z}^e_{\text{MOS}}, G)$
6 $\mathcal{E}_{\text{temp}}, \mathcal{Z}_{\text{temp}} \leftarrow \mathit{epsilonConstraint}(\mathcal{E}^e_{\text{MOS}}, \mathcal{Z}^e_{\text{MOS}}, G)$ // Second phase
7 $\mathcal{E}^w_{\text{MOS}}, \mathcal{Z}^V_{\text{MOS}} \leftarrow \mathit{removeNonMaximumCliques}(\mathcal{E}_{\text{temp}}, \mathcal{Z}_{\text{temp}})$

### 4.2. The two-phase strategy

The *Two-phase* strategy is designed to efficiently address the MOS problem by exploring the polynomial-time solution for finding the extreme supported points of MOS (see Section 3.4).

Algorithm 5 presents the pseudo-code for this strategy, which determines the set $\mathcal{Z}^V_{\text{MOS}}$ for a given graph $G$ through the execution of two distinct phases. In the initial phase, a dichotomic search based on weighted sum scalarizations computes the extreme supported points (Line 5). Subsequently, in the second phase, an $\varepsilon$-constraint approach is applied to identify the remaining weakly-nondominated points (Line 6).

The following sections provide a detailed explanation of both phases.

#### 4.2.1. Dichotomic weighted sum scalarization

In the first phase, the set of extreme supported points $\mathcal{Z}^e_{\text{MOS}}$ is computed using Algorithm 4.

To initiate this process, the lexicographical optimal points, $z^r$ and $z^s$, are initialized in the main Algorithm 5 (Lines 1 and 2). These points are initialized with values representing two known extreme supported points: $z^r = (0,0)$ corresponding to a "quasi-clique" with no vertices, which is optimal for minimizing the number of vertices (c.f. Remark 1); and $z^s = (|E|, -|V|)$ representing the entire graph $G$, which is optimal for maximizing the number of edges. Subsequently, the sets $\mathcal{Z}^e_{\text{MOS}}$ and $\mathcal{E}^e_{\text{MOS}}$ are initialized with these two points and their corresponding weakly-efficient quasi-cliques, respectively (see Lines 3 and 4 in Algorithm 5). Then, the dichotomic search procedure (Algorithm 4) is invoked to recursively find all the extreme supported points for MOS.

Unlike the baseline approach, which identifies exactly $|V| - \omega(G) + 1$ weakly-nondominated points, in the *Two-phase* strategy, the dichotomic search may find extreme-supported points corresponding to cliques that are not necessarily the maximum clique. One such example is the point $(0,0)$. Therefore, to ensure that the set $\mathcal{Z}^V_{\text{MOS}}$ contains exactly $|V| - \omega(G) + 1$ weakly-nondominated points, the procedure *removeNonMaximumCliques* is invoked in the main Algorithm 5 (Line 7) to remove those points representing non-maximum cliques. This is done by selecting all cliques found by the algorithm (i.e. the solutions with density 1.0), removing those with fewer vertices, and retaining the clique with maximum vertex cardinality.

#### 4.2.2. $\varepsilon$-constraint approach

In the second phase, an $\varepsilon$-constraint strategy is applied to identify all the remaining weakly-nondominated points not discovered in the first phase. To accomplish this, we employ an $\varepsilon$-constraint approach, executing $e$-DKS($G$, $k$), similar to the baseline approach presented in Algorithm 3.

Algorithm 6 shows the pseudo-code for this method that uses two temporary sets, $\mathcal{Z}_{\text{temp}}$ and $\mathcal{E}_{\text{temp}}$, to store the discovered weakly-nondominated points and their corresponding weakly-efficient quasi-cliques. Both $\mathcal{Z}_{\text{temp}}$ and $\mathcal{E}_{\text{temp}}$ are initialized with the set of extreme

**Algorithm 6:** epsilonConstraint

**input:** $\mathcal{E}^{e}_{\text{MOS}}$, $\mathcal{Z}^{e}_{\text{MOS}}$, $G$

1 $\mathcal{Z}_{\text{temp}} \leftarrow \mathcal{Z}^{e}_{\text{MOS}}$
2 $\mathcal{E}_{\text{temp}} \leftarrow \mathcal{E}^{e}_{\text{MOS}}$
3 Let $\mathcal{Z}_{\text{temp}} = \{z^1, \ldots, z^p\}$ such that $z^1_2 < z^2_2 < \cdots < z^p_2$
4 Let $z^i \in \arg\max_{z \in \mathcal{Z}} \left\{ z_1 \, : \, z^j_2 \neq z^i_2 + 1, \forall j > i, \text{with } i, j = 1, \ldots, p \right\}$
5 $k \leftarrow -z^i_2$
6 Let $G_{S^k} \in \{G_S \, : \, G_S \in \mathcal{E}_{\text{temp}}, |S| = k\}$
7 **while** $dens(G_{S^k}) < 1$ **do**
8  $k \leftarrow -z^i_2 - 1$
9  $G_{S^k} \leftarrow e\text{-}DKS(G, k)$
10  $\mathcal{E}_{\text{temp}} \leftarrow \mathcal{E}_{\text{temp}} \cup \{G_{S^k}\}$
11  $\mathcal{Z}_{\text{temp}} \leftarrow \mathcal{Z}_{\text{temp}} \cup \{(E(S^k), -|S^k|)\}$
12  Let $z^i \in \arg\max_{z \in \mathcal{Z}_{\text{temp}}} \{z_1 \, : \, -z_2 \leq k, z^j_2 \neq z^i_2 + 1, \forall j > i$, with $i, j = 1, \ldots, |\mathcal{Z}_{\text{temp}}|\}$
13 **end**
14 **return** $\mathcal{E}_{temp}, \mathcal{Z}_{temp}$

supported points found in the first phase and its corresponding set of quasi-cliques, respectively (Lines 1 and 2, respectively). This algorithm assumes that the points in $\mathcal{Z}_{\text{temp}}$ are sorted in decreasing order of the number of vertices (i.e., in increasing order of $z_2$-values), with $p = |\mathcal{Z}_{\text{temp}}|$ (Line 3). Next, the algorithm collects the maximum point $z^i \in \mathcal{Z}_{\text{temp}}$ such that the subsequent point, whose cardinality coordinate is $z^i_2 + 1$, was not yet identified (Line 4). Following this, the algorithm iteratively finds the weakly-nondominated point for each cardinality $k$ not yet discovered (Lines 7 to 13). As in our baseline approach, for $e$-DKS($G$, $k$) in Line 9, we adopt the M1 model (see formulation (2) in Section 4.1).

Algorithm 6 terminates once a maximum clique for $G$ is identified (see condition in Line 7).

In the *Two-phase* method, the runs of $e$-DKS($G$, $k$) are reduced in proportion to the number of extreme supported points discovered during the first phase.

### 4.3. The three-phase strategy

The *Three-phase* strategy (Algorithm 7) is proposed aiming to further reduce the runs of the $\varepsilon$-constraint problems $e$-DKS($G$, $k$). To accomplish this, this strategy comprises three distinct phases. In the initial phase, it employs the same dichotomic search as in the *Two-phase* method (refer to Section 4.2.1 and Algorithm 4). Subsequently, in the second phase, the algorithm applies a minimum-degree vertex-based local search to generate new candidates for weakly-efficient quasi-cliques (Line 6, Procedure $minD$). Finally, the third phase combines a maximum-degree vertex-based local search with an $\varepsilon$-constraint approach executing $e$-DKS($G$, $k$) to identify the remaining weakly-efficient quasi-cliques (Line 7, Procedure $maxD$).

In the following, we present both local search methods and explain how they are employed to generate new weakly-efficient quasi-cliques.

#### 4.3.1. Minimum degree vertex-based local search

This method leverages Propositions 7 and 8 to explore and discover nested quasi-cliques within existing ones using as a starting point each weakly-efficient quasi-clique corresponding to an extreme supported point identified in the previous phase.

Algorithm 8 presents the pseudo-code that uses two temporary sets, $\mathcal{Z}_{\text{temp}}$ and $\mathcal{E}_{\text{temp}}$, containing the extreme supported points and their corresponding weakly-efficient quasi-cliques, respectively, identified in the first phase (Lines 1 and 2, respectively). The points in $\mathcal{Z}_{\text{temp}}$ and

**Algorithm 7:** Three-phase strategy for MOS

**Data:** $G = (V, E)$
**Result:** $\mathcal{Z}^{V}_{\text{MOS}}$ and $\mathcal{E}^{w}_{\text{MOS}}$

1 $z^r \leftarrow (0, 0)$ // First phase
2 $z^s \leftarrow (|E|, -|V|)$
3 $\mathcal{Z}^{e}_{\text{MOS}} \leftarrow \{z^r, z^s\}$
4 $\mathcal{E}^{e}_{\text{MOS}} \leftarrow \{G_{S^r}, G\}$
5 $\mathcal{E}^{e}_{\text{MOS}}, \mathcal{Z}^{e}_{\text{MOS}} \leftarrow dichotomicSearch(z^r, z^s, \mathcal{E}^{e}_{\text{MOS}}, \mathcal{Z}^{e}_{\text{MOS}}, G)$
6 $\mathcal{E}_{\text{temp}}, \mathcal{Z}_{\text{temp}} \leftarrow minD(\mathcal{E}^{e}_{\text{MOS}}, \mathcal{Z}^{e}_{\text{MOS}})$ // Second phase
7 $\mathcal{E}_{\text{temp}}, \mathcal{Z}_{\text{temp}} \leftarrow maxD(\mathcal{E}_{\text{temp}}, \mathcal{Z}_{\text{temp}}, G)$ // Third phase
8 $\mathcal{E}^{w}_{\text{MOS}}, \mathcal{Z}^{V}_{\text{MOS}} \leftarrow removeNonMaximumCliques(\mathcal{E}_{\text{temp}}, \mathcal{Z}_{\text{temp}})$

**Algorithm 8:** minD method

**input:** $\mathcal{E}^{e}_{\text{MOS}}$, $\mathcal{Z}^{e}_{\text{MOS}}$

1 $\mathcal{Z}_{\text{temp}} \leftarrow \mathcal{Z}^{e}_{\text{MOS}}$
2 $\mathcal{E}_{\text{temp}} \leftarrow \mathcal{E}^{e}_{\text{MOS}}$
3 Let $\mathcal{Z}_{\text{temp}} = \{z^1, \ldots, z^p\}$ such that $z^1_2 < z^2_2 < \cdots < z^p_2$
4 Let $\mathcal{E}_{\text{temp}} = \{G_{S^1}, \ldots, G_{S^p}\}$ such that $z^i \in \mathcal{Z}_{\text{temp}}$ with $|E(S^i)| = z^i_1$ and $|S^i| = -z^i_2$, $\forall i = 1, \ldots, p$
5 $i \leftarrow 1$
6 $G_{S'} \leftarrow G_{S^i}$
7 **while** $i < p$ **and** $dens(G_{S'}) < 1$ **do**
8  $k \leftarrow -z^i_2 - 1$
9  **while** $k > -z^{i+1}_2$ **and** $dens(G_{S'}) < 1$ **do**
10   Let $v \in \arg\min_{v^* \in S'} \{deg_{G_{S'}}(v^*)\}$
11   $S' \leftarrow S' \setminus \{v\}$
12   **if** *weffTest*$(G_{S'}) = True$ **then**
13    $\mathcal{E}_{\text{temp}} \leftarrow \mathcal{E}_{\text{temp}} \cup \{G_{S'}\}$
14    $\mathcal{Z}_{\text{temp}} \leftarrow \mathcal{Z}_{\text{temp}} \cup \{(|E(S')|, -|S'|)\}$
15   **end**
16   $k \leftarrow k - 1$
17  **end**
18  **if** $dens(G_{S'}) < 1$ **then**
19   $i \leftarrow i + 1$
20   $G_{S'} \leftarrow G_{S^i}$
21  **end**
22 **end**
23 **return** $\mathcal{E}_{temp}, \mathcal{Z}_{temp}$

quasi-cliques in $\mathcal{E}_{\text{temp}}$ are assumed to be sorted in decreasing order of the number of vertices (i.e., in increasing order of $z_2$-values).

The method initiates with the weakly-efficient quasi-clique $G_{S'} \in \mathcal{E}_{\text{temp}}$ corresponding to the first extreme supported point $z^1 \in \mathcal{Z}_{\text{temp}}$ (Line 6). Subsequently, a new quasi-clique is iteratively generated by systematically removing from $G_{S'}$ the vertex with the minimum degree in $G_{S'}$ (Lines 10 and 11). This strategy reduces the vertex count while retaining as many edges as possible. The quasi-clique generation process is applied to yield new quasi-cliques across the entire interval from $z^i$ to $z^{i+1}$, with $1 \leq i < p$, for each sequential pair of extreme supported points, using the quasi-clique $G_{S^i}$ corresponding to $z^i$ as a starting point. To this end, in the external loop (Lines 7 to 22) the algorithm iterates over the extreme supported points, while in the internal loop (Lines 9 to 17) the algorithm iterates over each $k$ between the extreme supported points $z^i$ and $z^{i+1}$, with $k$ representing the vertex cardinality of the quasi-clique to be generated. When the extreme supported point $z^{i+1}$ is obtained (see the first condition in Line 9), then the algorithm restarts the generation process from its corresponding weakly-efficient quasi-clique $G_{S^{i+1}}$ (Line 20). In the vertex selection step (Line 10), ties are resolved by choosing the vertex with neighbours of smaller degrees. This strategic choice ensures that each iteration sets the stage for the next step, guaranteeing that $G_{S'}$ will feature a more suitable vertex for removal in the subsequent iteration.

The *weffTest* procedure (Line 12) evaluates the efficiency of each newly generated quasi-clique of size $k$. The resulting quasi-clique $G_{S'}$ is guaranteed to be weakly-efficient if one of the following conditions is met:

(*i*) The quasi-clique from which $G_{S'}$ was generated is weakly-efficient and the degree of the removed vertex is zero, meaning that $G_{S'}$ is the best possible quasi-clique of cardinality $k$.
(*ii*) $G_{S'}$ is a clique.
(*iii*) $G_{S'}$ is either a supported quasi-clique satisfying condition ($a$) or a non-supported quasi-clique satisfying condition ($b$) of Proposition 11, which will be presented subsequently.

Conditions ($i$), ($ii$), and ($iii$) are derived from Propositions 9, 10, and 11, respectively, which are presented below.

**Proposition 9.** *For a given graph $G = (V, E)$, let the subgraph $G_S$ induced by $S \subseteq V$ be a weakly-efficient quasi-clique for the MOS problem, and let $v \in S$ be a vertex such that $deg_{G_S}(v) = 0$. Let $S' = S \setminus \{v\}$. Then, $G_{S'}$ is a weakly-efficient quasi-clique.*

**Proof.** If there exists a quasi-clique $G_{S''}$ with $|S''| < |S'|$ and $|E(S'')| > |E(S')|$, then also $|S''| < |S|$ and $|E(S'')| > |E(S)| = |E(S')|$, contradicting the assumption. □

**Proposition 10.** *For a given graph $G = (V, E)$, let the subgraph $G_S$ induced by $S \subseteq V$ be a quasi-clique. If $dens(G_S) = 1$, then $G_S$ is an efficient quasi-clique for the MOS problem.*

**Proof.** This follows directly from the fact that, if $dens(G_S) = 1$, then there cannot be a graph with fewer nodes and the same number of edges, or with the same number of nodes and more edges. □

**Proposition 11.** *Let $\mathcal{Z}^e_{MOS} = \{z^1, z^2, \ldots, z^p\}$ such that $z^1_2 > z^2_2 > \cdots > z^p_2$ and let $z^i$ and $z^{i+1}$, with $1 \leq i < p$, be two adjacent extreme supported points. Let $w_1 = -z^{i+1}_2 + z^i_2$ and $w_2 = z^{i+1}_1 - z^i_1$. Let $\tilde{z} = (\tilde{z}_1, \tilde{z}_2)$ be a point that corresponds to a feasible quasi-clique $G_S$. Then, $G_S$ is weakly-efficient for the MOS problem if one of the following conditions is met:*

*(a)* $w_1 \cdot (\tilde{z}_1 - z^i_1) + w_2 \cdot (\tilde{z}_2 - z^i_2) = 0$ *($\tilde{z}$ is a non-extreme supported point).*
*(b)* $w_1 \cdot (\tilde{z}_1 + 1 - z^i_1) + w_2 \cdot (\tilde{z}_2 - z^i_2) > 0$ *($\tilde{z}$ is a non-supported point).*

**Proof.** First note that if $z^i$ and $z^{i+1}$ are two adjacent extreme supported points then $z^1$ and $z^2$ both correspond to optimal solutions of WS-MOS($G$, $w_1$, $w_2$) with the same weighted-sum objective value of $\bar{c} := w_1 \cdot z^i_1 + w_2 \cdot z^i_2 = w_1 \cdot z^{i+1}_1 + w_2 \cdot z^{i+1}_2$, where $w_1 > 0$ and $w_2 > 0$. Then, condition ($a$) implies that $w_1 \cdot \tilde{z}_1 + w_2 \cdot \tilde{z}_2 = w_1 \cdot z^i_1 + w_2 \cdot z^i_2 = \bar{c}$. In this case, $\tilde{z}$ is also optimal for WS-MOS($G$, $w_1$, $w_2$), and hence it corresponds to an efficient solution of the MOS problem. Similarly, condition ($b$) implies that $w_1 \cdot (\tilde{z}_1 + 1) + w_2 \cdot \tilde{z}_2 > \bar{c}$. By the optimality of the objective value of $\bar{c}$ for WS-MOS($G$, $w_1$, $w_2$), this implies that there cannot exist a quasi-clique $G_{S'}$ with $-|S'| \geq \tilde{z}_2$ and $|E(S')| > \tilde{z}_1$ that could potentially dominate $G_S$. It follows that $\tilde{z}$ corresponds to a weakly-efficient solution for the MOS problem. □

Note that to ensure a quasi-clique $G_S$ is efficient for the MOS problem, Proposition 11 must be revised. In such a case, $G_S$ must satisfy condition ($b$) and an additional condition stating that $w_1 \cdot (\tilde{z}_1 - z^i_1) + w_2 \cdot (\tilde{z}_2 + 1 - z^i_2) > 0$.

Following Algorithm 8, if the generated quasi-clique is proven to be weakly-efficient, it is added to the set $\mathcal{E}_{\text{temp}}$ (Line 13) and its corresponding weakly-nondominated point is added to the set $\mathcal{Z}_{\text{temp}}$ (Line 14).

Algorithm 8 terminates either when it finds a clique (i.e., $dens(G_{S'}) = 1$) or when it reaches the last extreme supported point $z^p$, meaning that one quasi-clique candidate has been generated for each value of $k$ in the interval between each sequential pair of extreme supported points in $\mathcal{Z}_{\text{temp}}$. At the end of *minD* both temporary sets, $\mathcal{Z}_{\text{temp}}$ storing the weakly-nondominated points identified so far and $\mathcal{E}_{\text{temp}}$ containing the respective weakly-efficient quasi-cliques, are returned to the main Algorithm 7 for use in the subsequent phase.

It is worth noticing that, in this approach, a new weakly-efficient quasi-clique may be generated from three sources: a weakly-efficient quasi-clique corresponding to an extreme-supported point, a weakly-efficient quasi-clique from the preceding iteration, or a quasi-clique generated in the preceding iteration and whose efficiency was not confirmed by the available sufficient conditions. The latter implies that the clique identified by *minD* does not necessarily correspond to the maximum clique. Therefore, similar to the *Two-phase strategy*, to guarantee that the set $\mathcal{Z}^V_{\text{MOS}}$ contains precisely $|V| - \omega(G) + 1$ weakly-nondominated points for a given graph $G = (V, E)$, the *Three-phase* strategy employs the *removeNonMaximumCliques* procedure within the main Algorithm 7 (Line 8) to eliminate those points representing non-maximum cliques that may have been generated by the dichotomic search in the first phase and by the *minD* local search in the second phase.

#### 4.3.2. Maximum degree vertex-based local search

In this section, we introduce the procedure *maxD*, a strategy that integrates local search with an $\varepsilon$-constraint approach to identify the remaining weakly-nondominated points not identified in the previous phases. Like *minD*, the *maxD* strategy generates new quasi-cliques from existing ones. However, this strategy consistently builds new quasi-cliques by adding a new vertex to a weakly-efficient quasi-clique.

Algorithm 9 outlines the pseudo-code for *maxD*, which takes as input the temporary sets $\mathcal{Z}_{\text{temp}}$ and $\mathcal{E}_{\text{temp}}$, as well as the graph $G = (V, E)$. At this point, these sets contain all the weakly-nondominated points and weakly-efficient quasi-cliques, respectively, discovered during the first and second phases. The points in $\mathcal{Z}_{\text{temp}}$ and quasi-cliques in $\mathcal{E}_{\text{temp}}$ are assumed to be sorted in increasing order of the number of vertices (i.e., in decreasing order of $z_2$-values). The algorithm begins by collecting the minimum point $z^i \in \mathcal{Z}_{\text{temp}}$ such that the subsequent point, with a cardinality coordinate of $z^i_2 - 1$, has not been identified yet (Lines 3 and 5). Next, starting with the weakly-efficient quasi-clique $G_{S'}$ corresponding to the collected point $z^i$ (Line 6), the algorithm iteratively generates a new quasi-clique by systematically selecting the vertex $v$ from $V \setminus S'$, where $v$ is the vertex with the maximum degree with respect to $G_{S'}$, and adding it to $G_{S'}$. This strategy minimally increases the vertex cardinality while adding as many edges as possible. In the vertex selection step (Line 7), ties are broken by choosing the vertex with the maximum degree in the entire graph $G$. This step allows the algorithm to explore and extend quasi-cliques in the subsequent iterations, as vertices with higher degrees tend to contribute to the formation of denser quasi-cliques. The generation process is executed to produce new quasi-cliques for each value of $k$ for which a weakly-efficient quasi-clique has not been discovered yet (Lines 4 to 16).

Similar to *minD* (Algorithm 8), in procedure *weffTest* (Line 9), the efficiency of each new quasi-clique is evaluated. The resulting quasi-clique $G_{S'}$ is guaranteed to be weakly-efficient if one of the following conditions is met:

(*i*) The degree of the added vertex $v$ is equal to the largest degree in $G$. This means that the best possible quasi-clique of cardinality $k$ was generated.
(*ii*) $G_{S'}$ is a clique.
(*iii*) $G_{S'}$ is either a supported quasi-clique satisfying condition ($a$) or a non-supported quasi-clique satisfying condition ($b$) of Proposition 11.

Note that conditions ($ii$) and ($iii$) are identical to those presented in the *minD* local search (see Section 4.3.1), while condition ($i$) is derived from Proposition 12 below.

**Algorithm 9:** maxD method

**input:** $\mathcal{E}_{\text{temp}}$, $\mathcal{Z}_{\text{temp}}$, and $G$

1 Let $\mathcal{Z}_{\text{temp}} = \{z^1, \dots, z^p\}$ such that $z_2^1 > z_2^2 > \cdots > z_2^p$

2 Let $\mathcal{E}_{\text{temp}} = \{G_{S^1}, \dots, G_{S^p}\}$ such that $z^i \in \mathcal{Z}_{\text{temp}}$ with $|E(S^i)| = z_1^i$ and $|S^i| = -z_2^i$, $\forall i = 1, \dots, p$

3 Let $Z = \underset{z \in \mathcal{Z}_{\text{temp}}}{\arg\min} \left\{ z_1 \,:\, z_2^j \neq z_2^i - 1, \forall j > i, \text{with } i, j = 1, \dots, p \right\}$

4 **while** $Z \neq \emptyset$ **do**

5   Let $z^i \in Z$

6   $G_{S'} \leftarrow G_{S^i}$

7   Let $v \in \underset{v^* \in V \setminus S'}{\arg\max} \{deg_{G_{S' \cup \{v^*\}}}(v^*)\}$

8   $S' \leftarrow S' \cup \{v\}$

9   **if** *weffTest*$(G_{S'}) = False$ **then**

10    $k \leftarrow |S'|$

11    $G_{S'} \leftarrow e\text{-}DKS(G, k)$

12   **end**

13   $\mathcal{Z}_{\text{temp}} \leftarrow \mathcal{Z}_{\text{temp}} \cup \{(|E(S')|, -|S'|)\}$

14   $\mathcal{E}_{\text{temp}} \leftarrow \mathcal{E}_{\text{temp}} \cup \{G_{S'}\}$

15   Let $Z = \underset{z \in \mathcal{Z}_{\text{temp}}}{\arg\min} \left\{ z_1 \,:\, -z_2 \geq k, z_2^j \neq z_2^i - 1, \forall j > i, \text{with } i, j = 1, \dots, |\mathcal{Z}_{\text{temp}}| \right\}$

16 **end**

17 **return** $\mathcal{E}_{temp}, \mathcal{Z}_{temp}$

**Proposition 12.** *For a given graph $G = (V, E)$, let the subgraph $G_S$ induced by $S \subseteq V$ be an efficient quasi-clique for the MOS problem. Let $v \in \arg\max_{v^* \in V \setminus S}\{deg_{G_{S\cup\{v^*\}}}(v^*)\}$ and $deg_{G_{S\cup\{v\}}}(v) = \Delta(G)$, with $\Delta(G) > 0$. Let $S' = S \cup \{v\}$. Then, the subgraph $G_{S'}$ induced by $S'$ is an efficient quasi-clique.*

**Proof.** The increment in terms of the number of edges from $G_S$ to $G_{S'}$ is the largest possible. □

If a quasi-clique is not proven to be weakly-efficient, then the algorithm turns to an $\epsilon$-constraint approach, executing $e$-DKS($G$, $k$) for the current value of $k$ (Line 11). This ensures that each iteration consistently produces a weakly-efficient quasi-clique, thereby enhancing the likelihood of generating new quasi-cliques that are efficient through local search in subsequent iterations. Similar to previous strategies (Algorithms 3 and 5), this method employs the M1 model (see formulation (2) in Section 4.1) for $e$-DKS($G$, $k$).

Algorithm 9 concludes its execution when there are no remaining nondominated points to be discovered (i.e., $Z = \emptyset$), as indicated by the condition in Line 4. This condition ensures that the algorithm has generated new weakly-efficient quasi-cliques for every $k$ value for which quasi-cliques have not been discovered in earlier phases. Upon finalization, the sets $\mathcal{E}_{\text{temp}}$ and $\mathcal{Z}_{\text{temp}}$ containing the weakly-efficient quasi-cliques and weakly-nondominated points, respectively, identified throughout all three phases are returned to the main Algorithm 7 (Line 17).

In the *Three-phase* strategy, the frequency of $e$-DKS($G$, $k$) runs is reduced in proportion to the number of weakly-nondominated points discovered during the first and the second phases, as well as the number of weakly-efficient quasi-cliques generated by the *maxD* local search.

## 5. Computational experiments

In the computational experiments, we use a set of real-life sparse graph instances obtained from the University of Florida Sparse Matrix Collection (Davis & Hu, 2011), along with the graph *Homer* from Trick (2002). All graphs were made undirected and simple by ignoring the direction of the arcs and removing self-loops and multiple edges. Although our experiments mainly focus on real-life graphs, we also test our approaches on synthetic graphs for illustrative purposes. To generate these graphs, we chose the extended Barabási-Albert model $G(n, m, p, q)$ (Albert & Barabási, 2000) due to its ability to produce graphs with properties similar to many real-world networks, such as a power-law degree distribution, network growth, and preferential attachment (see Barabási and Pósfai (2016) for details). In the model, $n$ is the number of vertices, $m$ is the number of edges with which a new node attaches to existing nodes, $p$ is the probability value for adding an edge between existing nodes, and $q$ is the probability value of rewiring existing edges. Since the majority of real-life graphs tend to exhibit sparse characteristics (Veremyev et al., 2016), we generated synthetic sparse graphs[1] with the following parameters: $n \in \{100, 200, 500, 1000, 2000\}$, $m \in \{2, 4, 6\}$, $p = 0.25$, and $q = 0.25$. Table 1 shows the number of vertices, number of edges, and density of the synthetic graphs in the first group and the real-life graphs in the second group.

**Table 1**
Characterization of the tested instances.

| Graph | Type | \|V\| | \|E\| | *dens* |
|---|---|---|---|---|
| n100m2 | $G(n, m, p, q)$ | 100 | 310 | 0.06 |
| n100m4 | $G(n, m, p, q)$ | 100 | 556 | 0.11 |
| n100m6 | $G(n, m, p, q)$ | 100 | 864 | 0.17 |
| n200m2 | $G(n, m, p, q)$ | 200 | 600 | 0.03 |
| n200m4 | $G(n, m, p, q)$ | 200 | 1 136 | 0.06 |
| n200m6 | $G(n, m, p, q)$ | 200 | 1 740 | 0.09 |
| n500m2 | $G(n, m, p, q)$ | 500 | 1 526 | 0.01 |
| n500m4 | $G(n, m, p, q)$ | 500 | 2 932 | 0.02 |
| n500m6 | $G(n, m, p, q)$ | 500 | 4 518 | 0.04 |
| n1000m5 | $G(n, m, p, q)$ | 1000 | 2 982 | <0.01 |
| n1000m4 | $G(n, m, p, q)$ | 1000 | 5 876 | 0.01 |
| n2000m2 | $G(n, m, p, q)$ | 2000 | 5 874 | <0.01 |
| polbooks | Copurchasing | 105 | 441 | 0.08 |
| smallW | Collaboration | 233 | 994 | 0.04 |
| USAir97 | Transportation | 332 | 2 126 | 0.04 |
| erdos971 | Collaboration | 433 | 1 314 | 0.01 |
| celegans-metabolic | Biological | 453 | 2 025 | 0.02 |
| harvard500 | Internet | 500 | 2 043 | 0.02 |
| homer | Book | 556 | 1 628 | 0.01 |
| email | Internet | 1133 | 5 451 | <0.01 |
| netscience | Collaboration | 1461 | 2 742 | <0.01 |
| yeast | Biological | 2284 | 6 646 | <0.01 |
| ca-GrQc | Collaboration | 5241 | 14 484 | <0.01 |
| erdos02 | Collaboration | 5534 | 8 472 | <0.01 |
| geom | Collaboration | 6158 | 11 898 | <0.01 |
| as-735 | Communication | 6474 | 12 572 | <0.01 |
| EVA | Ownership | 7253 | 6 711 | <0.01 |

For comparison purposes, we use the $e$-DKS-based $\epsilon$-constraint method (Algorithm 3) as a baseline against which we evaluate the performance of the *Two-phase* and the *Three*-phase strategies (Algorithms 5 and 7, respectively). Throughout this section, we will refer to the $e$-DKS-based $\epsilon$-constraint method approach as the *baseline* method.

The proposed algorithms were implemented in Python 3.8, and the models WS-MOS (formulation (1) in Section 3.4) and M1 (formulation (2) in Section 4.1) were solved using the Gurobi Optimizer version 10.0.2 with the Python interface. Multithreading was disabled in the MILP solver (thread count limit set to 1) because experimental tests showed that the M1 model performs better with this setting for most tested graphs. The computational experiments were conducted on a computer cluster with two Intel Xeon Silver 4210R 2.4G processors with 10 cores each and 251 GB of memory running under DebianGNU\Linux 12 (Bookworm).

Our goal with the *Two-phase* and *Three-phase* methods is to reduce the number of $\epsilon$-constraint runs, that is, to identify a large percentage of weakly-nondominated points without having to invoke $e$-DKS($G$, $k$) to solve the problem. Therefore, to assess our proposed approaches, we employ two metrics. Firstly, we consider the number of weakly-nondominated points found in each phase of the algorithms. This number is expected to be higher in phases not executing $\varepsilon$-constraint and relatively lower when this method is applied. The second metric involves evaluating the CPU time spent in solving the MOS problem by the proposed methods.

[1] Available at https://github.com/danielascherer/MOQC.

**Table 2**
Results for $e$-DKS-based $\varepsilon$-constraint, Two-phase and Three-phase approaches.

| Graph | $e$-DKS-based $\varepsilon$-constraint | | Two-phase | | | | Three-phase | | | | | | $\lvert\mathcal{Z}^V_{\text{MOS}}\rvert$ | $\lvert\mathcal{Z}_{\text{MOQC}}\rvert$ |
|---|---|---|---|---|---|---|---|---|---|---|---|---|---|---|
| | *#points* | *t* | *#points* | *%DS* | *%ε* | *t* | *#points* | *%DS* | *%minD* | *%maxD* | *%ε* | *t* | | |
| n100m2 | 95 | 2.3 | 96 | 13.5 | 86.5 | 2.3 | 96 | 13.5 | 71.9 | 0.0 | 14.6 | 1.9 | 95 | 95 |
| n100m4 | 92 | 5.5 | 93 | 14.0 | 86.0 | 5.4 | 93 | 14.0 | 69.9 | 0.0 | 16.1 | 4.6 | 92 | 92 |
| n100m6 | 92 | 158.7 | 93 | 18.3 | 81.7 | 157.4 | 95 | 17.9 | 48.4 | 1.1 | 32.6 | 155.2 | 92 | 92 |
| n200m2 | 196 | 12.4 | 197 | 7.6 | 92.4 | 12.2 | 198 | 7.6 | 74.7 | 2.0 | 15.7 | 10.7 | 196 | 196 |
| n200m4 | 193 | 38.4 | 194 | 11.3 | 88.7 | 38.1 | 195 | 11.3 | 69.7 | 0.5 | 18.5 | 35.1 | 193 | 193 |
| n200m6 | 193 | 2 017.1 | 194 | 12.9 | 87.1 | 2 016.7 | 195 | 12.8 | 51.3 | 1.0 | 34.9 | 2 012.6 | 193 | 193 |
| n500m2 | 496 | 114.3 | 497 | 5.8 | 94.2 | 113.6 | 497 | 5.8 | 77.5 | 1.2 | 15.5 | 97.7 | 496 | 496 |
| n500m4 | 494 | 565.9 | 495 | 8.7 | 91.3 | 562.2 | 497 | 8.7 | 75.1 | 3.2 | 13.1 | 542.1 | 494 | 494 |
| n500m6 | 491 | 3 092.1 | 492 | 9.1 | 90.9 | 3 090.3 | 492 | 9.1 | 73.4 | 2.8 | 14.6 | 3 020.4 | 491 | 491 |
| n1000m2 | 996 | 438.3 | 997 | 3.9 | 96.1 | 435.0 | 998 | 3.9 | 80.7 | 5.5 | 9.9 | 363.2 | 996 | 996 |
| n1000m4 | 993 | 1 512.4 | 994 | 6.7 | 93.3 | 1 501.9 | 994 | 6.7 | 75.6 | 4.4 | 13.3 | 1 331.7 | 993 | 993 |
| n2000m2 | 1 997 | 3 905.2 | 1 998 | 2.9 | 97.1 | 3 878.2 | 1 998 | 2.9 | 77.8 | 8.0 | 11.3 | 3 346.1 | 1 997 | 1997 |
| polbooks | 100 | 7.1 | 101 | 15.8 | 84.2 | 7.0 | 101 | 15.8 | 42.6 | 4.0 | 37.6 | 6.4 | 100 | 100 |
| smallW | 227 | 23.0 | 228 | 8.8 | 91.2 | 22.7 | 228 | 8.8 | 68.4 | 3.5 | 19.3 | 18.8 | 227 | 227 |
| USAir97 | 311 | 58.6 | 312 | 15.0 | 84.9 | 57.1 | 313 | 15.0 | 72.8 | 3.2 | 8.9 | 44.1 | 311 | 311 |
| erdos971 | 427 | 53.4 | 428 | 8.2 | 91.8 | 52.5 | 430 | 8.1 | 76.3 | 3.3 | 12.3 | 45.1 | 427 | 427 |
| celegans-metabolic | 445 | 53.8 | 446 | 10.5 | 89.5 | 52.1 | 446 | 10.5 | 74.9 | 6.5 | 8.1 | 31.6 | 445 | 445 |
| harvard500 | 480 | 442.1 | 481 | 7.3 | 92.7 | 437.4 | 481 | 7.3 | 66.1 | 2.7 | 23.9 | 427.2 | 480 | 480 |
| homer | 544 | 42.8 | 545 | 7.3 | 92.7 | 41.9 | 545 | 7.3 | 78.2 | 5.7 | 8.8 | 30.7 | 544 | 544 |
| e-mail | 1 122 | 17 609.1 | 1 123 | 4.7 | 95.3 | 17 569.4 | 1 123 | 4.7 | 63.3 | 5.0 | 27.0 | 17 432.9 | 1 122 | 1 122 |
| netscience | 1 442 | 382.0 | 1 443 | 3.6 | 96.4 | 375.8 | 1 443 | 3.6 | 43.0 | 38.4 | 15.0 | 253.2 | 1 442 | 1 442 |
| yeast | 2 276 | 1 334.1 | 2 277 | 2.7 | 97.3 | 1 304.6 | 2 280 | 2.7 | 78.0 | 5.5 | 13.8 | 877.2 | 2 276 | 2 276 |
| ca-GrQc | 5 198 | 20 408.7 | 5 199 | 2.6 | 97.4 | 20 201.8 | 5 199 | 2.6 | 62.3 | 22.5 | 12.6 | 16 298.5 | 5 198 | 5 198 |
| erdos02 | 5 528 | 1 212.1 | 5 529 | 0.9 | 99.1 | 1 206.0 | 5 529 | 0.9 | 95.6 | 1.5 | 2.0 | 342.2 | 5 528 | 5 528 |
| geom | 6 137 | 5 120.1 | 6 138 | 2.2 | 97.8 | 5 015.4 | 6 138 | 2.2 | 67.4 | 23.7 | 6.6 | 728.6 | 6 137 | 6 137 |
| as-735 | 6 465 | 3 139.6 | 6 466 | 1.0 | 99.0 | 3 122.0 | 6 466 | 1.0 | 93.8 | 3.6 | 1.6 | 203.3 | 6 465 | 6 465 |
| EVA | 7 250 | 3 808.5 | 7 251 | 0.7 | 99.3 | 3 771.8 | 7 252 | 0.7 | 86.9 | 9.2 | 3.3 | 216.2 | 7 250 | 7 250 |

Given the relative simplicity of the mapping process (Section 3.3), which involves merely computing the density for each point in $\mathcal{Z}^V_{\text{MOS}}$ and selecting the nondominated points for problem MOQC, the time spent on this process is considered insignificant and thus not included in the reported results.

In the following sections, we discuss the results obtained from our experiments.

### 5.1. Experimental results for MOS problem

Table 2 summarizes the results of using the *baseline* method, as well as the *Two-phase* and *Three-phase* strategies, to compute the set $\mathcal{Z}^V_{\text{MOS}}$ of weakly-nondominated points for the MOS problem across all the instances described in Table 1. Column *#points* indicates the total number of weakly-nondominated points discovered by each approach. The columns *%DS*, *%ε*, *%minD*, and *%maxD* show the percentage of the total weakly-nondominated points identified by dichotomic search based on weighted sum, $\varepsilon$-constraint, *minD* and *maxD* local search methods, respectively, during their respective phases and approaches. Column $\%t(\varepsilon)$ presents the proportion of time dedicated to the $\varepsilon$-constraint method within the *Three-phase* approach, expressed as a percentage of the total time in column *t*. Finally, columns $|\mathcal{Z}^V_{\text{MOS}}|$ and $|\mathcal{Z}_{\text{MOQC}}|$ indicate the final number of weakly-nondominated points of problem MOS and the number of nondominated points of problem MOQC for each tested instance.

#### 5.1.1. Number of weakly-nondominated points

Recall that throughout its execution, the *baseline* algorithm identifies consistently $|V| - \omega(G) + 1$ weakly-nondominated points for problem MOS. This corresponds precisely to the number of weakly-nondominated points in $\mathcal{Z}^V_{\text{MOS}}$, as evidenced by the matching values in Columns *#points* and $|\mathcal{Z}^V_{\text{MOS}}|$ in Table 2. This is because this algorithm stops identifying points once a maximum clique is discovered. In contrast, the *Two-phase* strategy may find more than $|V| - \omega(G) + 1$ weakly-nondominated points, as it depends on the number of extreme-supported points identified in the first phase by the dichotomic search algorithm. This observation is reflected in the values reported in Column *#points* for this approach. Across all tested graphs, the number of weakly-nondominated points identified by the *Two-phase* strategy exceeds $|\mathcal{Z}^V_{\text{MOS}}|$ by only one, which corresponds to the extreme-supported point $(0, 0)$.

In the context of the *Three-phase* approach, the number of weakly-nondominated points identified for problem MOS may also exceed $|V| - \omega(G) + 1$, as it relies on the discovery of weakly-nondominated points by dichotomic search and *minD* local search. However, our findings indicate that, for most graphs, this strategy produces results comparable to those of the baseline and *Two-phase* approaches. Apart from the difference related to the presence of the extreme-supported point $(0, 0)$, minor discrepancies are observed only for the synthetic graphs *n100m6*, *n200m2*, *n200m4*, *n200m6*, *n500m4*, and *n1000m2*, and for the real-life graphs *EVA*, *yeast*, *erdos971*, and *USAir97*. This indicates that despite the *minD* local search not always identifying the maximum quasi-clique for every graph $G$, this does not significantly affect the *Three-phase* strategy by creating too many unnecessary quasi-cliques.

*Extreme-supported points.* The experiments reveal the presence of supported points in all tested graphs. Regarding the extreme-supported ones, as reported in column *%DS*, the results show that among the synthetic graphs, those with fewer vertices, such as *n100m2*, *n100m3*, and *n100m4* exhibit higher percentages (13.5%, 14.0%, and 18.3%, respectively), while the percentages for the others range from 3.9% to 9.1%. For the real-life graphs, those such as *EVA*, *erdos02*, and *as-735* exhibit lower percentages of extreme-supported points (0.7%, 0.9%, and 1.0%, respectively), while graphs like *celegans-metabolic*, *USAir97*, and *polbooks* present higher percentages, reaching up to 10.5%, 15.0%, and 15.8%, respectively.

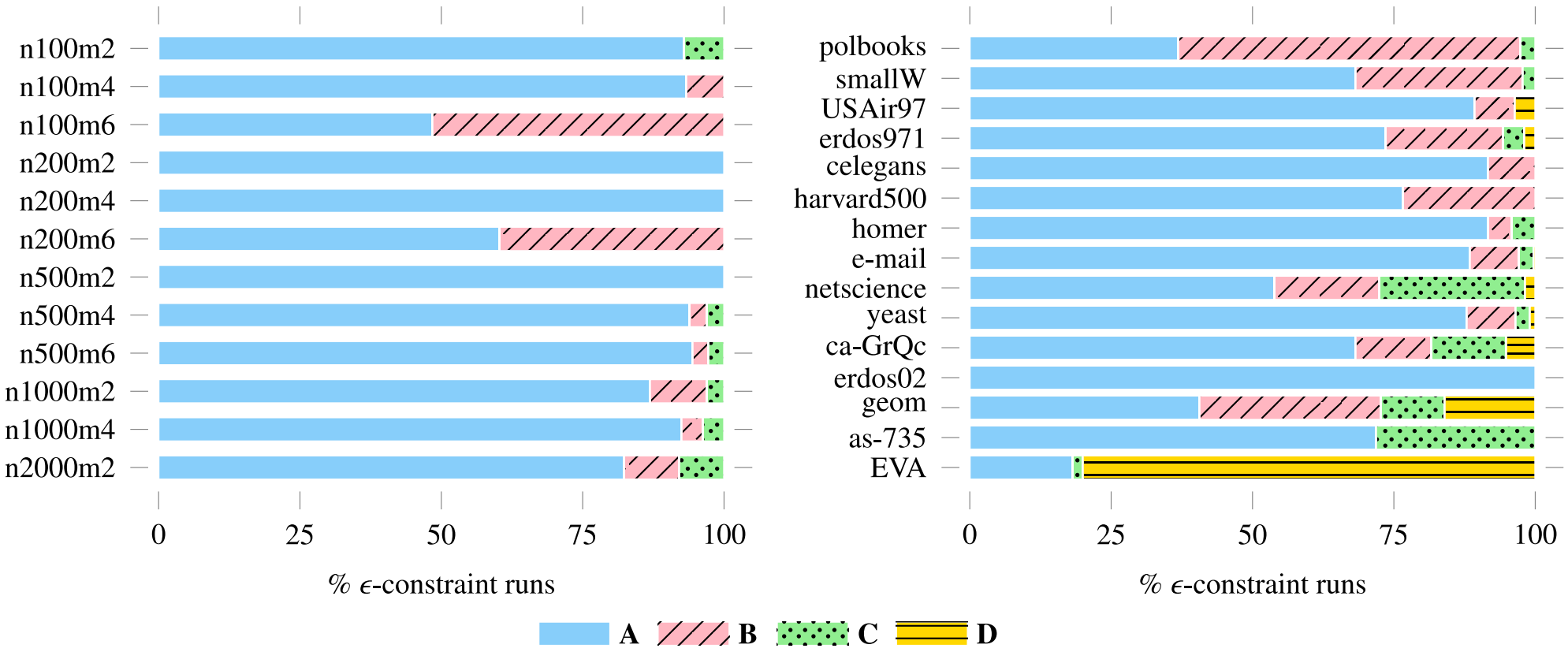

**Fig. 2.** Distribution of $\varepsilon$-constraint runs across four segments (A, B, C, and D) of each graph using the *Three-phase* strategy. Segments are defined as: (A) $\omega(G) \leq k \leq \frac{|V|}{4}$, (B) $\frac{|V|}{4} < k \leq \frac{|V|}{2}$, (C) $\frac{|V|}{2} < k \leq \frac{3\cdot|V|}{4}$, and (D) $\frac{3\cdot|V|}{4} < k \leq |V|$.

When dealing with graphs lacking supported points, both the *Two-phase* and *Three-phase* strategies are expected to face certain limitations. The *Two-phase* strategy is anticipated to perform as the baseline approach, identifying the weakly-efficient quasi-cliques for each cardinality value $k$ ranging from $|V|$ to $\omega(G)$ through the $\varepsilon$-constraint method. The *Three-phase* strategy is limited in this case to generating new quasi-cliques only between the two points $(|E|, -|V|)$ and $(0, 0)$ and does not leverage condition $a$ of Proposition 11 to assess the efficiency of the produced quasi-cliques. Consequently, it may identify only a limited number of weakly-nondominated points using the local search algorithms. However, it is noteworthy that our experiments conducted on real-life sparse graphs revealed the presence of supported points in all tested graphs. Therefore, the occurrence of graphs lacking such points may not be common in practical scenarios. However, further investigations are necessary to confirm this observation.

*5.1.2. Frequency of ε-constraint runs*

Concerning the frequency of runs of the $\varepsilon$-constraint problem, we analysed the results in columns $\%\varepsilon$. In the *baseline* approach, this percentage is 100% since all points in $\mathcal{Z}^{V}_{\text{MOS}}$ are identified by applying the $\varepsilon$-constraint method. However, for *Two-phase*, this result depends on the number of extreme-supported points found by the dichotomic search. Therefore, as expected, Table 2 shows a slight reduction in $\varepsilon$-constraint runs for graphs with fewer extreme-supported points, such as *n1000m2* and *n500m2* for the synthetic graphs and *EVA*, *erdos02*, and *as-735* for the real-life graphs, and a more significant decrease for those with more extreme-supported points (*n100m2*, *n100m4*, *n100m6*, *celegans-metabolic*, *USAir97*, and *polbooks*).

For the *Three-phase* strategy, the frequency of $\varepsilon$-constraint runs depends on the ability of both *minD* and *maxD* local searches to generate new *good* quasi-cliques, along with an effective mechanism to evaluate if the generated quasi-cliques are weakly-efficient. The results in columns *%minD* and *%maxD* indicate that generating new quasi-cliques from existing ones, using an extreme-supported point as a starting point, and applying Propositions 9, 10, 11, and 12 to assess the efficiency of the produced quasi-cliques is a quite effective strategy for identifying weakly-nondominated points for the MOS problem for both synthetic and real-life graphs. The *minD* method yields noteworthy outcomes, building quasi-cliques that are weakly-efficient and accounting for as much as 80.7% of the weakly-nondominated points for the synthetic graph *n1000m2* and 95.6% for the real-life graph *erdos02*. Even in less favourable cases, e.g., the synthetic graph *n100m6* and the real-life graphs *polbooks* and *netscience*, the algorithm builds 48.4%, 42.6% and 43.0% of the weakly-efficient quasi-cliques, respectively. The *maxD* local search complements the successful outcomes achieved by *minD* for all tested graphs, except for the synthetic graphs *n100m2* and *n100m4*. Noteworthy, in real-life graphs such as *netscience*, *geom*, and *ca-GrQc* this approach generates 38.4%, 23.8%, and 22.4%, respectively, of the weakly-efficient quasi-cliques. This contribution raises the percentage of the total number of weakly-efficient quasi-cliques produced for these graphs to 81.4%, 91.2%, and 84.7%, respectively.

Considering the higher percentage of weakly-nondominated points found by both *minD* and *maxD* local search methods, the *Three-phase* strategy demonstrates significant reductions in the frequency of $\varepsilon$-constraint runs across all tested graphs. Notably, instances such as the synthetic graph *n1000m2* and the real-life graphs *as-735*, *erdos02*, and *EVA* account for the highest reductions, with only 9.9%, 1.6%, 2.0%, and 3.3% of the weakly-nondominated points being identified through the $\varepsilon$-constraint approach, respectively. In less favourable scenarios, such as the synthetic graph *n100m6* and the real-life graphs *polbooks*, *e-mail*, and *harvard500*, the $\varepsilon$-constraint approach is needed to identify 32.6%, 37.6%, 27.0%, and 23.9%, respectively, of the weakly-nondominated points.

Aiming to identify specific regions of the graph where the $\varepsilon$-constraint method is most required to solve the problem, we partitioned the entire graph area into four segments. Segment A corresponds to $\omega(G) \leq k \leq \frac{|V|}{4}$, segment B to $\frac{|V|}{4} < k \leq \frac{|V|}{2}$, segment C to $\frac{|V|}{2} < k \leq \frac{3\cdot|V|}{4}$, and segment D to $\frac{3\cdot|V|}{4} < k \leq |V|$. Fig. 2 shows the distribution of $\varepsilon$-constraint runs as percentages across these segments for each tested graph. The figure highlights that in most graphs, the $\varepsilon$-constraint method is predominantly invoked to identify points located in segment A, that is, towards the direction of the maximum clique. As this pattern occurs regardless of the size or density of the graphs, it can be attributed to the inherent graph topology, where non-supported points, whose efficiency of their corresponding quasi-clique cannot be confirmed by Proposition 11, are more prevalent closer to the maximum clique. The graphs *polbooks* and *EVA* exhibited higher percentages of $\varepsilon$-constraint runs in segments B and D, respectively. For *EVA*, the *minD* and *maxD* local searches consistently failed to find weakly-efficient solutions in segment D whenever the $\varepsilon$-constraint was invoked. In contrast, for *polbooks*, while solutions were correctly identified 92.6% of the time in segment B, their efficiency could not be confirmed by the propositions. The outcomes related to the points validated by Propositions 9, 10, 11, and 12 are further discussed in Section 5.1.4.

Fig. 3 displays the results from the *Three-phase* strategy on the *Harvard500* graph, showing the set $\mathcal{Z}^{V}_{\text{MOS}}$ of weakly-nondominated points. Extreme-supported points identified by the dichotomic search based on weighted sum scalarization, are denoted by solid black circles. Weakly-nondominated points generated by the *minD* and *maxD* local search methods are marked with open blue circles, while those discovered through the $\varepsilon$-constraint runs are indicated with crossed red circles. The figure is divided into parts for clarity. Fig. 3(a) shows all weakly-nondominated points found by the strategy. Fig. 3(b) zooms into the subset of weakly-nondominated points on the right side of Fig. 3(a), corresponding to segment A in Fig. 2. This subset spans from the point $(210, -21)$, representing the maximum clique, to $(821, -95)$. Fig. 3(c) focuses on the points located on the left side of Fig. 3(a), covering the range from the point $(1877, -376)$ to the point $(2043, -500)$, corresponding to the segment D in Fig. 2. Notice that all points in this segment were discovered via dichotomic search or local searches, eliminating the need for the $\varepsilon$-constraint runs.

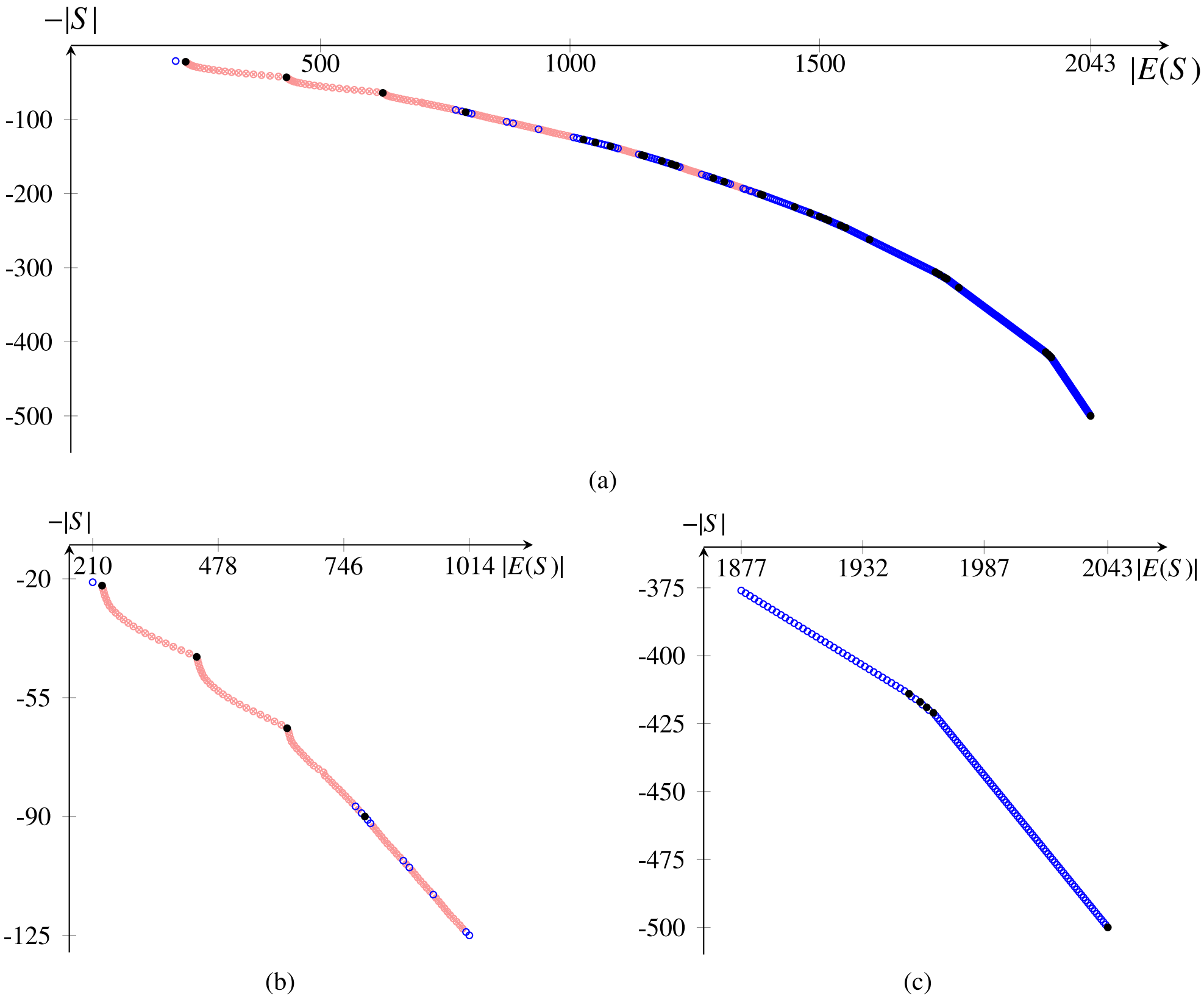

**Fig. 3.** Weakly-nondominated points in $\mathcal{Z}^V_{\text{MOS}}$ identified by the *Three-phase* strategy for *Harward500*, highlighting extreme-supported points (solid black circles), weakly-nondominated points identified by *minD* and *maxD* (open blue circles) and by $\varepsilon$-constraint (crossed red circles). ($a$) Entire weakly-nondominated set. ($b$) Subset of points ranging from $(210, -21)$ to $(1014, -125)$, and ($c$) from $(1877, -376)$ to $(2043, -500)$. (For interpretation of the references to colour in this figure legend, the reader is referred to the web version of this article.)

Additionally, we investigated how the frequency of $\varepsilon$-constraint runs relates to various graph properties, such as density, size, and edge count. Our analysis reveals that the number of $\varepsilon$-constraint runs varies with these features, suggesting no consistent relationships.

#### 5.1.3. Running time

Fig. 4 displays the time spent by the baseline approach across the same four segments A, B, C, and D defined in Section 5.1.2. Based on these results, except for graph *polbooks*, segment *A* is the most time-consuming region for all the tested graphs. This segment also corresponds to the highest number of $\varepsilon$-constraint runs, as depicted in Fig. 2.

The results related to the running time in seconds spent by each approach are presented in column *t* (Table 2).

For both the *baseline* and *Two-phase* methods, the reported run time corresponds to the total elapsed time recorded by the Gurobi solver for identifying the set of weakly-nondominated points $\mathcal{Z}^V_{\text{MOS}}$. Specifically, for *Two-phase*, this duration includes the time spent on finding extreme-supported points during the first phase, as well as the time dedicated to discovering the remaining points in the second phase. The *Two-phase* strategy demonstrates a minor reduction in running time compared to the baseline approach. This slight reduction may be attributed to Gurobi's strategy of initially solving the relaxation of model M1 (referenced in formulation (2)) when addressing the *e*-DKS problems in the baseline method. Consequently, it can identify the extreme-supported points within a timeframe comparable to employing the linear programming relaxation of the weighted-sum scalarization, WS-MOS (referenced in formulation (1)), during the dichotomic search.

For the *Three-phase* strategy, the reported running time corresponds to the total time spent across the three phases to identify set $\mathcal{Z}^V_{\text{MOS}}$. The results reported in Table 2 indicate that for graphs such as *EVA*, *as-735*, *geom*, and *erdos02*, the total time spent by this approach exhibits a significant reduction compared to the *Two-phase* method: 94.3%, 93.5%, 85.5%, and 71.6%, respectively. For the remaining tested graphs, including the synthetic ones, the observed reduction in percentage ranges from 0.2% to 39.4%.

In less favourable cases, specifically with the *n200m6*, *n100m6*, *n500m6*, and *n500m4* for synthetic graphs, and *e-mail* and *Harvard500* for real-life graphs, the *Three-phase* strategy achieves only a slight reduction in running time, ranging from 0.2% to 3.6%. This minimal time savings is due to a combination of factors: these graphs are the most time-demanding for solving the e-DKS problems in segment A, requiring over 94% of the total time, as illustrated in Fig. 4, and the *Three-phase* strategy identifies only a few points in this segment. A more significant time reduction could be achieved if the *Three-phase* strategy could identify more non-supported points in the first segment. In fact, the *minD* and *maxD* local search methods effectively generate a large percentage of weakly-efficient quasi-cliques in this segment, ranging from 41% to 100%. However, their efficiency cannot be confirmed by the available propositions. In the following section, we discuss the outcomes related to the points validated by each proposition.

#### 5.1.4. Analysing the strength of optimality conditions

This section presents the results of applying Propositions 9, 10, 11, and 12 as sufficient conditions to evaluate the (weak) efficiency of quasi-cliques generated by the local search approaches within the *Three-phase* strategy. In Table 3, the columns *%P9*, *%P10*, *%P11.a*, *%P11.b*, and *%P12* indicate the percentage of quasi-cliques successfully validated by each proposition (9, 10, 11.*a*, 11.*b*, and 12, respectively) for each tested graph, relative to the total weakly-efficient quasi-cliques identified by the *minD* and *maxD* local searches methods. Column *%Unproven* shows the percentage of quasi-cliques that, despite being weakly-efficient, could not be proven as such by the propositions.

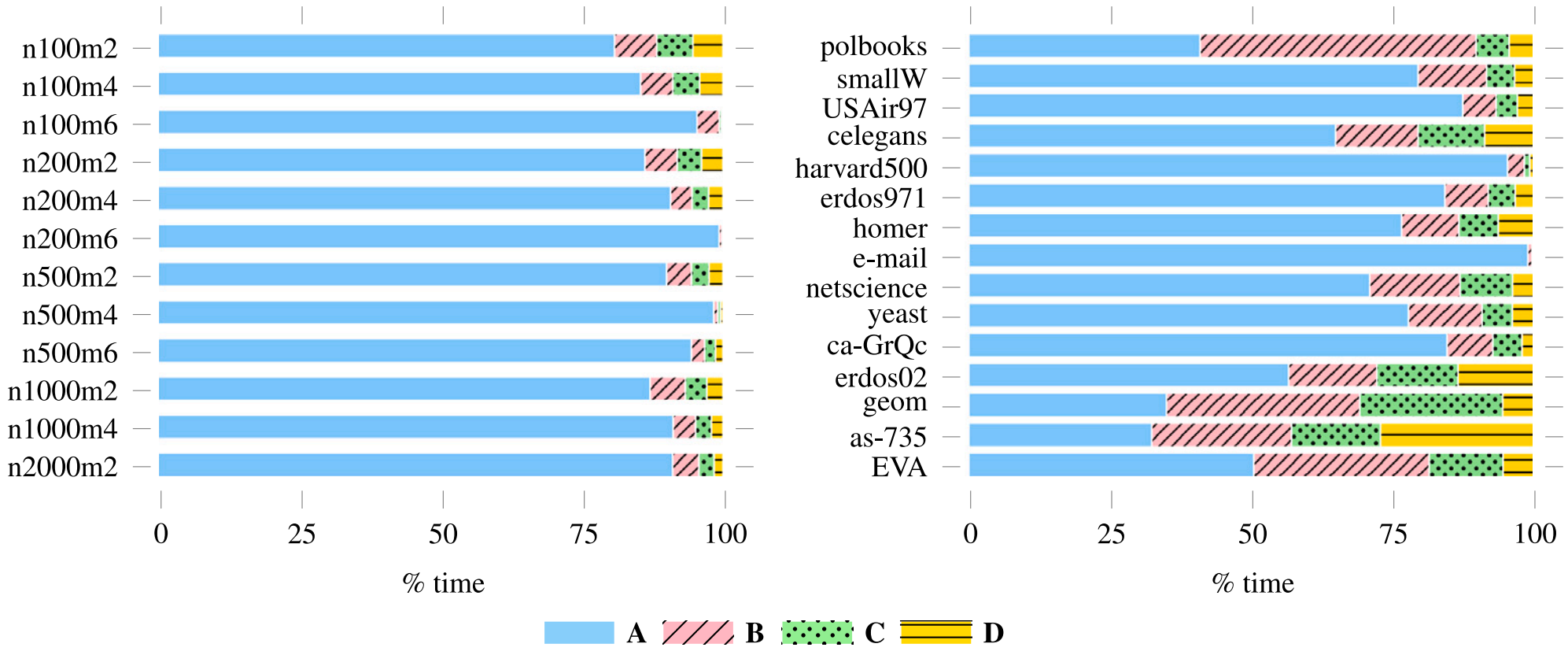

**Fig. 4.** Distribution of the time spent by the *baseline* approach across four segments (A, B, C, and D) of each graph. Segments are defined as: (A) $\omega(G) \leq k \leq \frac{|V|}{4}$, (B) $\frac{|V|}{4} < k \leq \frac{|V|}{2}$, (C) $\frac{|V|}{2} < k \leq \frac{3\cdot|V|}{4}$, and (D) $\frac{3\cdot|V|}{4} < k \leq |V|$.

Our findings indicate that for the synthetic graphs, Proposition 9 did not confirm the efficiency of any quasi-cliques, meaning that the local search process does not lead to quasi-cliques with vertices of degree zero. For the real-life graphs, this proposition effectively confirmed weak efficiency for 46.7% of the tested graphs. The most significant results are observed for the *geom* and *netscience* graphs, with 9.17% and 9.03% of the weakly-efficient quasi-cliques validated by this proposition, respectively.

Given that the *minD* local search stops the quasi-cliques generation upon identifying a clique, as expected, the percentage of validations by Proposition 10 across all tested graphs is minimal, not exceeding 2 validations. Specifically, for the *netscience* graph, no validation occurred under this proposition because the point representing the maximum clique was already identified as an extreme-supported point in the first phase through dichotomic search (Algorithm 4).

Concerning Proposition 11.*a*, validation percentages are 25.53% for the *polbooks* graph and 48.89% for the *netscience* graph. For the remaining graphs, validation rates range from 53.19% for *n100m6* to 94.38% for *erdos02*. The results for this proposition primarily depend on two factors: the existence of supported points, which is influenced by the topology of the graph, and the effectiveness of the *minD* and *maxD* local search methods in generating weakly-efficient quasi-cliques.

The percentages of weakly-efficient quasi-cliques validated by Proposition 11.*b* are generally below 50%, ranging from 5.61% to 42.55%, with *polbooks* being an exception at 72.34%. It is crucial to note that Proposition 11.*b* targets those non-supported points near the boundary of $\mathcal{Z}^{\leq}_{\text{MOS}}$. Such proximity is significant enough that their corresponding quasi-cliques can be conclusively identified as weakly-efficient. Consequently, similar to Proposition 11.*a*, the validation percentage relies on the occurrence of these particular non-supported points, which are influenced by the inherent topology of the graphs, and on the efficacy of the local search methods.

No weakly-efficient quasi-clique was validated by Proposition 12 because the tested graphs, such as *n200m4*, *yeast*, and *polbooks*, have at most two vertices with the maximum degree. This reflects the typical characteristic of both the real-life and generated synthetic graphs, which have few high-degree vertices and many low-degree vertices. Consequently, when the *maxD* local search adds a vertex to a weakly-efficient quasi-clique, the added vertex's degree is unlikely to match the graph's maximum degree.

**Table 3**
Percentage of weakly-efficient quasi-cliques validated by Propositions 9, 10, 11.*a*, 11.*b*, and 12, in the *Three-phase* strategy.

| Graph | %P9 | %P10 | %P11.*a* | %P11.*b* | %P12 | %Unproven |
|---|---|---|---|---|---|---|
| n100m2 | 0.00 | 1.45 | 85.51 | 13.04 | 0.00 | 100.00 |
| n100m4 | 0.00 | 1.54 | 86.15 | 12.31 | 0.00 | 100.00 |
| n100m6 | 0.00 | 4.26 | 53.19 | 42.55 | 0.00 | 93.55 |
| n200m2 | 0.00 | 0.66 | 86.18 | 13.16 | 0.00 | 93.55 |
| n200m4 | 0.00 | 1.46 | 82.48 | 16.06 | 0.00 | 100.00 |
| n200m6 | 0.00 | 0.98 | 72.55 | 26.47 | 0.00 | 95.59 |
| n500m2 | 0.00 | 0.26 | 79.80 | 19.95 | 0.00 | 93.51 |
| n500m4 | 0.00 | 0.26 | 69.41 | 30.33 | 0.00 | 86.15 |
| n500m6 | 0.00 | 0.27 | 76.27 | 23.47 | 0.00 | 97.22 |
| n1000m2 | 0.00 | 0.12 | 76.63 | 23.26 | 0.00 | 73.74 |
| n1000m4 | 0.00 | 0.13 | 67.80 | 32.08 | 0.00 | 77.27 |
| n2000m2 | 0.00 | 0.06 | 76.78 | 23.16 | 0.00 | 65.49 |
| polbooks | 0.00 | 2.13 | 25.53 | 72.34 | 0.00 | 94.74 |
| smallW | 0.00 | 0.61 | 84.76 | 14.63 | 0.00 | 97.73 |
| USAir97 | 0.00 | 0.84 | 76.89 | 22.27 | 0.00 | 89.29 |
| erdos971 | 0.29 | 0.29 | 71.93 | 27.49 | 0.00 | 66.04 |
| celegans-metabolic | 0.00 | 0.28 | 60.61 | 39.12 | 0.00 | 86.11 |
| harvard500 | 0.00 | 0.30 | 79.15 | 20.54 | 0.00 | 96.52 |
| homer | 0.66 | 0.22 | 78.77 | 20.35 | 0.00 | 83.33 |
| e-mail | 0.00 | 0.13 | 78.36 | 21.51 | 0.00 | 85.48 |
| netscience | 9.03 | 0.00 | 48.89 | 42.08 | 0.00 | 33.64 |
| yeast | 0.84 | 0.10 | 80.84 | 18.22 | 0.00 | 77.39 |
| ca-GrQc | 3.99 | 0.02 | 61.37 | 34.61 | 0.00 | 58.78 |
| erdos02 | 0.00 | 0.02 | 94.38 | 5.61 | 0.00 | 65.74 |
| geom | 9.17 | 0.02 | 56.08 | 34.74 | 0.00 | 11.33 |
| as-735 | 0.00 | 0.02 | 93.06 | 6.92 | 0.00 | 31.07 |
| EVA | 7.22 | 0.01 | 61.88 | 30.89 | 0.00 | 2.12 |

The results in the *%Unproven* column show a significant portion of weakly-efficient quasi-cliques generated by *minD* and *maxD* whose weakly efficiency could not be confirmed by the available propositions. For synthetic graphs, this portion reaches 100% in *n100m2*, *n100m4*, and *n200m4*, while for real-life graphs, it is as high as 97.73% in *smallW*. Notably, except for *EVA*, for all the other synthetic and real-life graphs, most of these unproven quasi-cliques are located in the segment with the highest percentage of $\epsilon$-constraint runs, which are segment B for *polbooks* and segment A for the other graphs as illustrated in Fig. 2. These outcomes suggest that, by excluding the $\epsilon$-constraint from its third phase, the *Three-phase* strategy emerges as a promising heuristic approach for addressing the MOQC problem.

### 5.2. Results for MOQC problem

The number of nondominated points for the MOQC problem as obtained by the mapping process outlined in Section 3.3 are presented in column $|\mathcal{Z}_{\text{MOQC}}|$ in Table 2.

In every tested graph, the weakly-nondominated points in $\mathcal{Z}^V_{\text{MOS}}$ identified for the MOS problem directly correspond to the nondominated points for the MOQC problem.

An interesting result from the analysis of all tested graphs is the absence of supported points for the MOQC problem, except for those corresponding to the entire graph $(dens(G), |V|)$ and the maximum clique. This lack of supported points reinforces the value of addressing MOQC through the MOS problem.

## 6. Conclusion

In this paper, we introduced and systematically investigated the MOQC problem, which is a novel perspective on identifying quasi-cliques in simple undirected graphs by simultaneously maximizing vertex cardinality and edge density. This perspective addresses the limitations inherent in single-objective variants of quasi-clique problems, such as the MQC and DKS problems, by eliminating the need for prior or precise preference information about cardinality and density. To efficiently tackle this problem, our methodology is based on exploring the relations among MOQC, its single-objective counterparts (MQC and DKS), and a pivotal bi-objective optimization problem, the MOS problem. This exploration elucidates several key properties of both the MOQC and MOS problems and, alongside analysis of inherent properties of quasi-cliques, provides a comprehensive solution strategy for identifying efficient quasi-cliques for the MOQC problem.

We introduced and detailed three exact solution strategies: a baseline approach based on the $\varepsilon$-constraint method; a *Two-phase* strategy that employs dichotomic search to identify the extreme-supported points and $\varepsilon$-constraint scalarization to identify the remaining weakly-nondominated points; and a *Three-phase* method that combines the aforementioned dichotomic search with local search based on vertex degree information, together with $\varepsilon$-constraint scalarization. While all three strategies proved effective for the MOQC problem, the *Three-phase* method distinguished itself by its superior performance in terms of running time and its efficiency in reducing the frequency of calls to execute the $\varepsilon$-constraint. The ability of the *Three-phase* strategy to discover new weakly-efficient quasi-cliques, supported by an application of local search techniques and a mechanism for quasi-clique efficiency assessment, represents a significant step forward in addressing the MOQC problem.

Regarding the limitations of the proposed approaches, there are a few points to be mentioned. Firstly, the *Two-* and *Three-phase* strategies are not effective in solving the MOS problem for graphs that do not present extreme-supported points, as discussed in Section 5.1.1. Secondly, although both the *baseline* and *Two-phase* approaches can be easily parallelized, adapting the *Three-phase* algorithm to parallelization is challenging. This is due to the *minD* local search procedure, which relies on the solution of size $k$ to generate the solution of size $k-1$ between two extreme-supported points. Similarly, in the *maxD* local search, the solution of size $k$ is required to generate the solution of size $k+1$.

Looking ahead, our work opens several avenues for future research. A potential direction would be to consider pruning techniques such as *k-core* or *k-truss* decomposition as preprocessing steps to extend the applicability of our methods to larger-scale real-life instances. This would include a careful investigation into whether the structure of the original graph is preserved such that the optimal solutions after pruning remain consistent with those of the original graph.

Additionally, investigate whether the *Three-phase* strategy maintains its time efficiency when the *baseline* and *Two-phase* approaches execute under different parallelization conditions. In this same direction, a *Four-phase* strategy could be considered to enhance parallelization. This approach would split the third phase of the *Three-phase* method into two independent phases: the first part using a *maxD* local search, and the second part applying the $\varepsilon$-constraint method to discover points not identified in the preceding phases. This would allow parallelization of both the first and fourth phases.

Another avenue could involve approaching the MOQC problem with implicit enumeration methods or other exact strategies. Further exploration might focus on developing more sophisticated local search algorithms, specifically designed to generate weakly-efficient quasi-cliques during the second and third phases of the *Three-phase* strategy. Investigating additional conditions to ensure the efficiency of the generated quasi-cliques also represents another promising area of exploration. Exploring ways to address graphs lacking supported points for the MOS problem also emerges as an important consideration. Additionally, there is potential for developing heuristic methods suited for larger-scale graphs.

Another opportunity for future work involves adapting our approaches to guarantee that the discovered quasi-cliques are connected. For the baseline and *Two-phase* approaches, this could be easily achieved by leveraging our recent contribution (dos Santos et al., 2024), which proposes adding constraints to well-known MILP formulations for DKS and MQC problems, such as the M1 model (see formulation (2)) used in our current work, to ensure the connectedness property of quasi-cliques. For the *Three-phase* strategy, further investigation is required to ensure that the *minD* and *maxD* local search methods produce weakly-efficient quasi-cliques that are guaranteed to be connected. Ensuring connectedness could potentially broaden the applicability of our strategies to a wide range of real-world scenarios.

Incorporating other relevant objectives into the MOQC problem, such as connectivity, is also a compelling direction for future work. In such a case, the MOQC problem would maximize density, cardinality, and connectivity simultaneously. This allows the identification of quasi-cliques that are not only dense and large but also robust in terms of connectivity, that is, not easy to disconnect.

In conclusion, this work not only introduces the MOQC problem to the academic community but also contributes by offering theoretical insights into both the MOQC and MOS problems and providing practical solutions for solving them. This effort represents a significant milestone in the field of quasi-clique optimization problems, opening new avenues for further research and application opportunities.

## CRediT authorship contribution statement

**Daniela Scherer dos Santos:** Writing – review & editing, Writing – original draft, Visualization, Validation, Software, Methodology, Investigation, Funding acquisition, Formal analysis, Data curation, Conceptualization. **Kathrin Klamroth:** Writing – review & editing, Visualization, Validation, Methodology, Investigation, Formal analysis, Data curation, Conceptualization. **Pedro Martins:** Writing – review & editing, Visualization, Validation, Supervision, Project administration, Methodology, Investigation, Formal analysis, Data curation, Conceptualization. **Luís Paquete:** Writing – review & editing, Writing – original draft, Visualization, Validation, Supervision, Resources, Project administration, Methodology, Investigation, Funding acquisition, Formal analysis, Data curation, Conceptualization.

## Acknowledgements

Daniela Scherer dos Santos acknowledges the Foundation for Science and Technology (FCT) for the Ph.D. fellowship, Portugal 2022.12082.BD. This work is partially funded by the FCT, I.P./MCTES through national funds (PIDDAC), within the scope of CISUC R&D Unit - UID/CEC/00326/2020.